\documentclass[pra,twocolumn,floatfix]{revtex4}
\usepackage{graphicx}
\usepackage{color}
\usepackage{amsmath}
\DeclareUnicodeCharacter{2009}{\,}
  
\begin{document}

\title{Rapidly-rotating quantum droplets confined in a harmonic potential}

\author{S. Nikolaou$^1$, G. M. Kavoulakis$^{1,2}$, and M. \"{O}gren$^{2,3}$}
\affiliation{$^1$Department of Mechanical Engineering, Hellenic Mediterranean University, P.O. Box 1939, 71004, Heraklion, Greece
\\
$^2$HMU Research Center, Institute of Emerging Technologies, 71004, Heraklion, 
Greece
\\
$^3$School of Science and Technology, \"{O}rebro University, 70182 \"{O}rebro, Sweden}
\date{\today}

\begin{abstract}

We consider a ``symmetric" quantum droplet in two spatial dimensions, which 
rotates in a harmonic potential, focusing mostly on the limit of ``rapid" rotation. 
We examine this problem using a purely numerical approach, as well as a semi-analytic 
Wigner-Seitz approximation (first developed by Baym, Pethick \textit{et al.}) for the description of the state with a vortex lattice. 
Within this approximation we assume that each vortex occupies a cylindrical cell, 
with the vortex-core size treated as a variational parameter. Working with a fixed 
angular momentum, as the angular momentum increases and depending on the atom number,
the droplet accommodates none, few, or many vortices, before it turns to center-of-mass 
excitation. For the case of a ``large" droplet, working with a fixed rotational frequency 
of the trap $\Omega$, as $\Omega$ approaches the trap frequency $\omega$, a vortex lattice 
forms, the number of vortices increases, the mean spacing between them decreases, while 
the ``size" of each vortex increases as compared to the size of each cell. In contrast to 
the well-known problem of contact interactions, where we have melting of the vortex lattice 
and highly-correlated many-body states, here no melting of the vortex lattice is present, 
even when $\Omega = \omega$. This difference is due to the fact that the droplet is 
self-bound. Actually, for $\Omega = \omega$, the ``smoothed" density distribution becomes 
a flat top, very much like the static droplet. When $\Omega$ exceeds $\omega$, the droplet 
maintains its shape and escapes to infinity, via center-of-mass motion. 

\end{abstract}

\maketitle

\section{Introduction}

In recent years the problem of quantum droplets has attracted a lot of attention. 
As Petrov \cite{Petrov} pointed out, quantum droplets are self-bound states, which 
may appear in binary mixtures of Bose-Einstein condensed atoms. Under typical 
conditions, the mean-field energy is the dominant part of the energy, while the 
(beyond-mean-field) corrections \cite{LHY} are very small. This is due to the 
diluteness condition -- which holds in the vast majority of experiments in 
(single-component) cold-atomic systems. In a two-component system, however, if 
we tune the inter-species and intra-species coupling constants, the mean-field 
energy may take any -- even an infinitesimally-small -- value. In this case, the 
energy due to the beyond-mean-field effects \cite{LHY} is no longer negligible, 
but rather it may balance the one due to the mean-field. Under these conditions 
the self-bound quantum droplets form. 

The literature on this problem is rather extensive. Here we refer to just a few of
the studies on quantum droplets, see, e.g., the review articles \cite{rrev1, rrev2}, 
and Refs.\,\cite{PA, th0, th1, th2, th3, th4, th5, th6, th7, th8, th9, th10, th11, 
th12, th14, th15, th16, EK, add1, th166, NKO, add2, add3, add4, add5, 
add6}. Experimentally, quantum droplets have been observed in two-component 
Bose-Einstein condensed gases \cite{qd7, qd8, qd8a, gd8b, qd8c}, but also in
single-component gases with strong dipolar interactions \cite{qd1, qd2, qd3, qd4, 
qd5, qd6}.

Quantum droplets offer a new system for studying the effects which are associated 
with ``superfluidity" \cite{Leggett}. The present study focuses on the rotational 
properties of this novel superfluid. Remarkably, since quantum droplets are self-bound, 
the nonlinear term that results from the interactions is partly attractive and partly 
repulsive. This is in sharp contrast to the case of contact interactions, where the 
scattering length has a fixed sign and as a result the effective interaction is either 
(purely) repulsive, or (purely) attractive. While the existence of these droplets does 
not require the presence of any trapping potential, the combination of an external 
potential with the nonlinear term gives rise to a very rich phase diagram. As a result, 
studying the rotational properties of quantum droplets under the action of an external 
potential is an interesting problem.

In the case of a harmonically-trapped single-component atomic condensate with an 
attractive contact interaction, the angular momentum is carried via center-of-mass 
excitation \cite{WGS, BM, PP}. On the other hand, for repulsive contact interactions, 
as the rotational frequency of the trap $\Omega$ increases, vortices enter the cloud 
and eventually a vortex lattice forms \cite{rev1, rev2, rev3, rev4, rev5}. As $\Omega$ 
increases even further, various interesting effects show up and the lattice ``melts". 
As $\Omega \to \omega^-$, where $\omega$ is the trap frequency, the system enters a 
highly-correlated regime, where the many-body state is given by a Laughlin-like state. 
This is due to the fact that as $\Omega$ increases, the effective potential, i.e., the 
sum of the confining potential, plus the centrifugal, becomes less and less steep, and 
it vanishes when $\Omega = \omega$.

The rotational properties of harmonically-trapped droplets have already been studied in 
several interesting studies. Reference \cite{th10} investigated the lowest-energy state
of the system for some fixed value of the total angular momentum $L \hbar$ and 
demonstrated the formation of vortices as $L$ increases. In Ref.\,\cite{EK} the same 
problem was considered. There, it was shown that depending on the atom number $N$, the 
frequency of the trapping potential $\omega$ and the total angular momentum $L \hbar$, 
various phases may appear. These include center-of-mass excitation, ghost vortices, 
as well as vortices of single and multiple quantization. In Ref.\,\cite{NKO} we worked 
also at fixed $L$ and found that as $L$ increases, there is a mixed state, where the 
droplet carries vortices and also undergoes center-of-mass excitation. Reference \cite{add2} 
considered the problem of a ``large" droplet, as $\Omega$ varies from ``small" values, 
up to the case $\Omega \to \omega$. The properties of the resulting vortex lattice were 
studied. Finally, Ref.\,\cite{add3} examined the case $\Omega = \omega$ and the limit 
of weak interactions and showed the formation of a triangular vortex lattice.  
   
An important observation is that there are two separate length scales, namely the 
size of the droplet $R$ (which is an increasing function of $N$) and the oscillator 
length $a_{\rm osc}$. For a fixed value of $a_{\rm osc}$, when the number of atoms $N$ 
which constitute the droplet is sufficiently small, the size of the droplet may be also 
much smaller than the oscillator length, i.e., $R \ll a_{\rm osc}$. In this case the 
droplet carries its angular momentum purely via excitation of its center of mass. As $N$ 
increases, the droplet size $R$ increases too, and as a result the droplet starts to 
get ``squeezed" by the harmonic potential. Eventually, when $R$ becomes comparable to 
$a_{\rm osc}$, instead of center-of-mass excitation, we have vortex excitation. 
Vortices start to penetrate the droplet, while for a sufficiently large value of $R$ and 
$\Omega$, a vortex lattice forms. However, as mentioned also above, we have shown in a 
recent study \cite{NKO} that for a sufficiently large value of $L$ -- or, equivalently, 
as $\Omega$ approaches $\omega$ -- the droplet, starts to undergo center-of-mass 
excitation. This is a ``mixed" state, where the droplet undergoes center-of-mass 
excitation, while it carries vortices. 

In the derived results which are presented below it is crucial that the relative 
coordinates separate from the center-of-mass coordinate \cite{PP}. This is true only 
in the case of harmonic confinement. Therefore, there are two completely independent 
and fully decoupled forms of excitation.  As a result, the droplet may carry its angular 
momentum via vortex excitation, via center-of-mass excitation, or via a superposition of 
these two kinds of excitation, depending on the value of the chosen parameters.

Two are the basic goals of the present study. The first one is to examine how these 
two independent kinds of excitation (i.e., center-of-mass and vortex) show up, depending
on the atom number and on the value of the angular momentum. The second goal is to 
investigate the properties of the vortex lattice, especially when $\Omega$ approaches 
$\omega$. This question includes the overall size of the droplet, as well as the ``size" 
of each vortex core. These questions are examined as $N$ and $\Omega$ are varied. We rely
on purely numerical results and mostly on a semi-analytic approach, using a (variational) 
Wigner-Seitz approximation for the vortex lattice, which has been used in the past \cite{GB1, 
GB2, GB3, GB4, GB5, GB6}. The main idea is that, in the presence of a vortex lattice there 
is a smoothed, slowly-varying, density distribution, and on top of that, a rapidly-varying 
density distribution due to the presence of the vortices. Treating the size of each vortex 
core variationally, we manage to develop a theory for the smoothed density distribution.  

In what follows below we first describe in Sec.\,II the model that we adopt. We consider 
a ``symmetric" droplet, where the two components have an equal population of atoms and 
also the coupling constants between the same components are equal to each other. We  
consider rotation of the droplet in the presence of a harmonic potential in purely two 
dimensions. In this section we start with the general model and also introduce our 
dimensionless quantities, giving also some experimentally-relevant scales. Then, we 
consider the limit of ``rapid" rotation, turning to the Wigner-Seitz approximation that 
we adopt, which is suitable for the description of a vortex lattice. There, we present 
the equations for the smoothed density distribution. In Sec.\,III we consider the 
Thomas-Fermi limit of the Wigner-Seitz approximation, for $\Omega = \omega$, deriving 
some analytic results. In Secs.\,IV and V we present the main results of our 
study. In Sec.\,IV we examine the first question that was mentioned in the previous 
paragraph, namely how the angular momentum is distributed between vortex and center-of-mass 
excitation. In Sec.\,V we examine the second set of questions, namely the properties 
of the vortex lattice as $N$ and $\Omega$ are varied. Finally, in Sec.\,VI we give a 
summary and a discussion of the main results of our study.

\section{Model}

\subsection{General equations}

First of all, we assume that there is a very tight potential along our $z$ axis,
which forces the atoms to move on the $xy$ plane and therefore we consider strictly 
two-dimensional motion. We also consider a two-component Bose-Einstein condensate,
where the two components, which we denote as ``$\uparrow$" and ``$\downarrow$", 
have an equal population of atoms, $N_\uparrow = N_\downarrow$, and also we consider 
equal masses $M$ for the two components. Regarding the atom-atom interactions, 
we assume an equal coupling between the same species, denoted as $g$, i.e., 
$g_{\uparrow \uparrow} = g_{\downarrow \downarrow} = g$, while the coupling 
between $\uparrow$ and $\downarrow$ is denoted as $g_{\uparrow \downarrow}$. 

Because of the assumptions mentioned above, there is a common order parameter for
the two components, which satisfies the following equation \cite{PA}
\begin{align}
  i \hbar \frac {\partial \Psi} {\partial t} = - \frac {\hbar^2} {2 M} {\boldsymbol \nabla}^2 \Psi  
  +  \frac 1 2 M \omega^2 r^2 \Psi +&
  \nonumber \\
  + \frac {4 \pi \hbar^2} {M \ln^2(a_{\uparrow \downarrow}/a)} |\Psi|^2 
  \ln \frac {|\Psi|^2} {2 {\sqrt e} n_0} \Psi.&
  \label{eq1}
\end{align}
Here $a$ and $a_{\uparrow \downarrow}$ are the two-dimensional scattering lengths 
for elastic atom-atom collisions between the same species (assumed to be equal for 
the two components) and for different species, respectively. Furthermore,
\begin{equation}
   n_0 = \frac {e^{-2 \gamma - 3/2}} {2 \pi} \frac {\ln(a_{\uparrow \downarrow}/a)} 
   {a a_{\uparrow \downarrow}}.
\end{equation}
Here $\gamma$ is Euler's constant, $\gamma \approx 0.5772$ and
\begin{align}
    \ln (a_{\uparrow \downarrow}/a) = \sqrt{\frac {\pi} 2} 
    \left( \frac {a_z} {a^{\rm 3D}} - \frac {a_z} {a_{\uparrow \downarrow}^{\rm 3D}} \right).
\end{align}
In the above equation $a_z$ is the ``width" of the droplet along the axis of rotation, 
and $a^{\rm 3D}$, $a_{\uparrow \downarrow}^{\rm 3D}$ are the three-dimensional scattering 
lengths for elastic atom-atom collisions between the same and different species, 
respectively. Introducing 
\begin{align}
  \Psi_0^2 = 2 \sqrt{e} n_0 = \frac {e^{-2 \gamma - 1}} {\pi} 
  \frac {\ln(a_{\uparrow \downarrow}/a)} {a a_{\uparrow \downarrow}},
\end{align}
and setting ${\tilde \Psi} = \Psi/\Psi_0$, Eq.\,(\ref{eq1}) becomes
\begin{align}
  i \frac {\partial {\tilde \Psi}} {\partial \tilde{t}} = 
  - \frac {1} {2} {\tilde {\boldsymbol \nabla}}^2 {\tilde \Psi}  
  +  \frac 1 2 {\tilde \omega}^2 {\tilde r}^2 {\tilde \Psi} 
  + |{\tilde \Psi}|^2 \ln |{\tilde \Psi}|^2 {\tilde \Psi}.
  \label{eq2}
\end{align}
Here $\tilde{t} = t/t_0$, where 
\begin{equation}
 t_0 = \frac {M a a_{\uparrow \downarrow} \ln(a_{\uparrow \downarrow}/a)} {4 \hbar e^{-2 \gamma - 1}}. 
\end{equation}
Also, ${\tilde r} = r/x_0$ and ${\tilde {\boldsymbol \nabla}}^2$ is the dimensionless Laplacian, 
with the unit of length being $x_0$, where
\begin{equation}
 x_0 = \sqrt{\frac {a a_{\uparrow \downarrow} \ln(a_{\uparrow \downarrow}/a)} 
 {4 e^{-2 \gamma - 1}}}.
\end{equation}
Furthermore, ${\tilde \omega} = \omega/\omega_0$, where the units of the frequency, $\omega_0$ 
and of the energy, $E_0$, are
\begin{equation}
  E_0 = \hbar \omega_0 = \frac {\hbar} {t_0} 
  = \frac {\hbar^2} {M x_0^2} = \frac {\hbar^2} {M a a_{\uparrow \downarrow}} 
  \frac {4 e^{-2 \gamma - 1}} {\ln(a_{\uparrow \downarrow}/a)}.
\end{equation}
The normalization condition takes the form
\begin{align}
 \int |{\tilde \Psi}|^2 \, d^2 {\tilde r} = \frac N {N_0},
\end{align}
where 
\begin{equation}
  N_0 = \Psi_0^2 x_0^2 = \frac 1 {4 \pi} \ln^2(a_{\uparrow \downarrow}/a),
\label{n0}
\end{equation}
which is the unit of $N$. 

Finally, the time-independent equation that corresponds to Eq.\,(\ref{eq2}) is derived after 
we set $\Psi({\tilde{\bf r}}, {\tilde t}) = \Psi({\tilde{\bf r}}) e^{- i {\tilde \mu} 
\tilde{t}}$, where ${\tilde \mu}$ is the dimensionless chemical potential, thus getting
\begin{align}
  - \frac {1} {2} {\tilde {\boldsymbol \nabla}}^2 {\tilde \Psi} 
  + \frac 1 2 {\tilde \omega}^2 {\tilde r}^2 {\tilde \Psi} 
  +  |{\tilde \Psi}|^2 \ln |{\tilde \Psi}|^2 {\tilde \Psi} = {\tilde{\mu}} {\tilde \Psi}.
  \label{eq22}
\end{align}
We stress that the ``tilde" used in the symbols in the present section, which represents 
dimensionless quantities, is dropped in all the equations which follow below, for convenience.

Equation (\ref{n0}) allows us to evaluate the actual (total) number of atoms in a droplet. 
For a typical value of $a_z = 0.1$ $\mu$m and $a^{\rm 3D} = 10.1$ nm, $a_{\uparrow 
\downarrow}^{\rm 3D} = -10.0$ nm, $\ln (a_{\uparrow \downarrow}/a) \approx 25$. Then, 
according to Eq.\,(\ref{n0}), $N_0 \approx 50$. Also, the unit of length $x_0$ turns out 
to be on the order of 1 $\mu$m. Finally, typical values of the two-dimensional density are 
$\approx 10^9$ ${\rm cm}^{-2}$, of the three-dimensional density are $10^{13}$ ${\rm cm}^{-3}$, 
$t_0$ is on the order of msec and the typical value of the trapping potential is hundreds of Hz.

The extended energy functional that we consider is, in dimensionless units, \cite{PA, th166} 
 \begin{align}
  {\cal E}(\Psi, \Psi^*) = E - L \Omega - \mu N =&
  \nonumber \\
  = \int \left( \frac {1} {2} |{\boldsymbol \nabla} \Psi|^2 
  + \frac 1 2 \omega^2 r^2 |\Psi|^2
  + \frac 1 2 |\Psi|^4 \ln \frac {|\Psi|^2} {\sqrt{e}} \right) \, d {\bf r}&
 \nonumber \\
  - \mu \int \Psi^* \Psi \, d {\bf r} - \Omega \int \Psi^* {\hat L} \Psi \, d {\bf r}.&
\label{funncc}
\end{align}
In the above equation $E$ is the total energy and $\mu$ is the chemical potential. Also, 
$\Psi$ is normalized to the scaled number of atoms, $\int |\Psi|^2 \, d {\bf r} = N$. The operator 
${\hat L}$ is that of the total angular momentum. We work with a fixed atom number and therefore 
$\mu$ is a Lagrange multiplier. Also, in some of the calculations that follow below we work 
with a fixed $L$ (in which case $\Omega$ is a Lagrange multiplier) \cite{GO}, and in other 
calculations we work with a fixed $\Omega$, in which case ${\cal E}(\Psi, \Psi^*)$ may be 
viewed as the energy of the system in the rotating frame. 

\subsection{Wigner-Seitz approximation}

Following Refs.\,\cite{GB1, GB2, GB3, GB4, GB5, GB6} we develop below an approximate method 
that allows us to study this problem in the presence of a vortex lattice. The assumptions 
which we make here are the following. First of all, we consider a droplet with a large atom 
number, as well as a large number of vortices. Also, we assume that the length scale over 
which the ``smoothed" density of the droplet changes (see also below) is much larger than the 
spatial size of each vortex core, or, equivalently, that the smoothed density does not change 
significantly over the core size. Actually, when $\Omega$ approaches $\omega$ -- which is 
the limit that we are mostly interested in -- the density of the droplet flattens out. As 
a result, the only significant variation of the smoothed density takes place solely at the edge 
of the droplet and it is constant elsewhere, as we see below.

Let us write the order parameter as
\begin{equation}
  \Psi({\bf r}) = \Phi({\bf r}) \cdot f({\bf r}) e^{i \phi({\bf r})},
  \label{1eq}
\end{equation}
and employ a Wigner-Seitz approximation. Since we have a triangular vortex lattice \cite{add2,add3},
we replace the triangular-shaped cells of the vortex lattice by cylindrical cells of equal 
radius $\ell_{\rm cell}$. Here $\Psi({\bf r})$ is the product of a real and slowly-varying 
envelope function, $\Phi({\bf r})$, times a rapidly varying factor, $f({\bf r})$, which 
vanishes at each vortex core, and has a phase $\phi$ which wraps by $2 \pi$ around each 
vortex. If we choose $f^2$ to average to unity over each unit cell of the lattice, 
\begin{align}
  \frac 1 {\pi \ell_{\rm cell}^2} \int_{\rm cell} f^2 \, d{\bf r}_c = 1,
\end{align}
where the integration is performed over one cell, then $\Phi^2({\bf r}) = n_s({\bf r})$ 
is the smoothed density profile of the droplet. Indeed, if we write the position vector 
${\bf r} = {\bf R}_j + {\boldsymbol \rho}_c$, where ${\bf R}_j$ is the center of the cell 
with index $j$, and ${\boldsymbol \rho}_c$ is the radial coordinate measured from each 
vortex line,
\begin{align}
   \int |\Psi({\bf r})|^2 \, d{\bf r} = \sum_j n_s({\bf R}_j) \int_{\rm cell} f^2 \, d{\bf r}_c =& \nonumber \\
  = \pi \ell^2_{\rm cell} \sum_j n_s({\bf R}_j).&
\end{align}
Converting the sum over cells into an integral,
\begin{equation}
 \pi \ell_{\rm cell}^2 \sum_j n_s({\bf R}_j) \to \int n_s({\bf r}) \, d{\bf r},
\end{equation}
we find that
\begin{equation}
  \int |\Psi({\bf r})|^2 \, d{\bf r} = \int n_s({\bf r}) \, d{\bf r} = N,
\end{equation}
where the integration is over all the $xy$ plane.

From Eq.\,(\ref{1eq}) it follows that the kinetic energy $K$ is
\begin{align}
 K = \frac 1 2 \int |{\boldsymbol \nabla} \Psi|^2 \, d {\bf r} =&
 \nonumber \\
 = \frac 1 2 \int [ ({\boldsymbol \nabla} \Phi)^2 + n_s f^2 ({\boldsymbol \nabla} \phi)^2
 + n_s ({\boldsymbol \nabla} f)^2) ] \,  d {\bf r},&
 \label{cc1}
\end{align}
where in the first term on the right we have assumed that $\Phi^2({\bf r})$ varies slowly 
across a unit cell of the vortex lattice and thus $f^2$ has been replaced by its average, 
i.e., unity. In Eq.\,(\ref{cc1}) there is also an integral which is proportional to
$\int {\boldsymbol \nabla}(f^2) \cdot {\boldsymbol \nabla} n_s \, d{\bf r}$. Integrating 
by parts and replacing $f^2$ by its average, we get an integral which is proportional to 
$\int {\boldsymbol \nabla}^2 n_s \, d{\bf r}$. From the divergence theorem this is equal 
to a surface integral, which vanishes.

We write the local velocity in some cell with index $j$ as the sum of a term that comes 
from solid body-rotation ${\bf \Omega} \times {\bf R}_j$, plus the local velocity around 
the vortex, ${\boldsymbol \nabla} \chi_j$, 
\begin{equation}
   {\boldsymbol \nabla} \phi = {\bf \Omega} \times {\bf R}_j + {\boldsymbol \nabla} \chi_j.
\label{cc2}
\end{equation}
Although the rotational velocity of the lattice is not necessarily equal to the rotational 
velocity of the trap, for a large system the difference between them is small \cite{GB3}. 

Combining Eqs.\,(\ref{cc1}) and (\ref{cc2}),
\begin{align}
 K = \int \frac 1 2 ({\boldsymbol \nabla} \Phi)^2 d \, {\bf r} 
 + \pi \ell_{\rm cell}^2 \sum_j \frac 1 2 n_s({\bf R}_j) \Omega^2 R_j^2 +&
 \nonumber \\
 + \sum_j n_s({\bf R}_j) \int_{\rm cell} \frac 1 2 \left[ ({\boldsymbol \nabla} f)^2 
 + f^2 ({\boldsymbol \nabla} \chi_j)^2 \right] \, d {\bf r}_c.&
 \label{cc3}
\end{align}
Here, the last integration is over each cell. Also, we have assumed that the cross term, 
which involves the inner product $({\bf \Omega} \times {\bf R}_j) \cdot {\boldsymbol \nabla} 
\chi_j$ is negligible, since the density does not vary significantly across the cell. 

Regarding the energy due to the harmonic potential, $V = (1/2) \omega^2 r^2$, its 
contribution to the energy may be written in the form
\begin{align}
  V = \frac 1 2 \omega^2 \int \Phi^2 f^2 r^2 \, d{\bf r} =&
  \nonumber \\
  = \sum_j \frac 1 2 n_s({\bf R}_j) \omega^2 
  \left( \pi \ell_{\rm cell}^2 R_j^2 + \int_{\rm cell} f^2 \rho_c^2 \, d{\bf r}_c\right).&
\label{poteq}
\end{align}

From Eqs.\,(\ref{cc3}) and (\ref{poteq}) we get that
\begin{align}
 K + V = \int \frac 1 2 ({\boldsymbol \nabla} \Phi)^2 d \, {\bf r} +&
 \nonumber \\
 + \pi \ell_{\rm cell}^2 \sum_j \frac 1 2 n_s({\bf R}_j) (\Omega^2 + \omega^2) R_j^2 +&
 \nonumber \\
 + \sum_j n_s({\bf R}_j) \int_{\rm cell} \frac 1 2 \left[ ({\boldsymbol \nabla} f)^2) 
 + f^2 ({\boldsymbol \nabla} \chi_j)^2 + \omega^2 f^2 \rho_c^2 \right] \, d {\bf r}_c.&
 \nonumber \\
\label{kpv}
\end{align}
In the neighbourhood of a given vortex, the local velocity ${\boldsymbol \nabla} \chi_j$ 
is $\approx {\boldsymbol {\hat \phi}}/\rho_c$, where $\boldsymbol{{\hat \phi}}$ is the unit 
vector around each cell. Also, $f$ is approximately radially symmetric about each vortex line. 
Thus, the first two terms in the integral in the last sum in Eq.\,(\ref{kpv}) take the 
form 
\begin{align}
  \int_{\rm cell} \frac 1 2 [({\boldsymbol \nabla} f)^2) + f^2 ({\boldsymbol \nabla} \chi_j)^2] 
 \,  d {\bf r}_c \approx&
\nonumber \\ 
 \approx \int_{\rm cell} \frac 1 2 \left[ \left( \frac {\partial f} {\partial \rho_c} 
 \right)^2 + \frac {f^2} {\rho_c^2} \right] \, d {\bf r}_c \equiv \pi a_j.&
\label{ke4}
\end{align}
Similarly, for the third term in the integral which appears in the last sum of Eq.\,(\ref{kpv}), 
we introduce the mean value of $\rho_c^2$ in cell with index $j$, $\langle \rho_{c,j}^2 \rangle$,
\begin{align}
 \langle \rho_{c,j}^2 \rangle =  
 \frac 1 {\pi \ell^2_{\rm cell}} \int_{\rm cell} f^2 \rho_c^2  \, d {\bf r}_c 
 \equiv \ell_{\rm cell}^2 b_j = \frac {b_j} {\Omega}.
\label{ke5}
\end{align}
Here we have used Feynman's formula for the vortex density, $n_v = \Omega/\pi = 1/(\pi 
\ell^2_{\rm cell})$. We assume that the quantities $a_j$ and $b_j$ are independent of the cell, 
and in what follows we will assume that all of them have a common value $a$, and $b$, 
respectively. This assumption implies that our approach is more accurate when $\Omega$
approaches $\omega$, as we discuss in more detail below.

Converting the two terms in Eq.\,(\ref{kpv}) which correspond to the ones in Eq.\,(\ref{ke4}) 
into an integral (over the whole $xy$ plane), 
\begin{align}
\sum_j n_s({\bf R}_j) \int_{\rm cell} \frac 1 2 \left[ ({\boldsymbol \nabla} f)^2) 
 + f^2 ({\boldsymbol \nabla} \chi_j)^2 \right] \, d {\bf r}_c& 
 \nonumber \\
 \to \frac a {\ell_{\rm cell}^2} \int n_s({\bf r}) \, d{\bf r} = a \Omega \int n_s({\bf r}) \, d{\bf r}.&
 \label{ffeq}
\end{align}
Similarly, the last term in Eq.\,(\ref{kpv}) may also be converted into an integral, which is given 
by 
\begin{align}
\sum_j n_s({\bf R}_j) \int_{\rm cell} \frac 1 2 \omega^2 f^2 \rho_c^2 \, d {\bf r}_c&
 \nonumber \\
 \to   \frac 1 2 \omega^2 \langle \rho_{c}^2 \rangle \int n_s({\bf r}) \, d{\bf r} 
   = \frac b 2 \frac {\omega^2} {\Omega} \int n_s({\bf r}) \, d{\bf r}.&
\label{ffeq2}
\end{align}

Finally, we examine the energy due to the nonlinear term, which is
\begin{align}
  \int \frac 1 2 |\Psi|^4 \ln \frac {|\Psi|^2} {\sqrt{e}} \, d {\bf r} =&
  \nonumber \\
 = \frac 1 2 \sum_j \int_{\rm cell} n_s^2({\bf R}_j) f^4(\rho_c) 
  \ln \frac {n_s({\bf R}_j) f^2(\rho_c)} {\sqrt{e}} \, d{\bf r}_c.&
\end{align}
To evaluate the integral over the unit cell we make the following ansatz for $f(\rho_c)$,
\begin{equation}
 f(\rho_c) =
 \begin{cases}
  (1 - \zeta/2)^{-1/2} (\rho_c/\xi) \,\, {\rm for} \,\, 0 \le \rho_c \le \xi, \,\, \\
  (1 - \zeta/2)^{-1/2} \,\, {\rm for} \,\, \xi < \rho_c \le \ell_{\rm cell},
   \end{cases}
\label{ansatz}
\end{equation}
where $\xi$ is a variational parameter and $\zeta = (\xi/\ell_{\rm cell})^2 \le 1$ is the 
fractional area of the vortex core in the unit cell. Therefore,
\begin{align}
\int \frac 1 2 |\Psi|^4 \ln \frac {|\Psi|^2} {\sqrt{e}} \, d {\bf r} =&
  \nonumber \\
  = \int \frac 1 2 \frac {\Phi^4({\bf r})} {(1 - \zeta/2)^2} 
  \left[ \left( 1 - \frac 2 3 \zeta \right)
   \ln \frac {\Phi^2({\bf r})} {(1 - \zeta/2) \sqrt{e}} - \frac {\zeta} 9 \right] \, d {\bf r}.&
   \nonumber \\
\label{inte}
\end{align}
To get the last equality we converted the sum into an integral.

Therefore, from Eqs.\,(\ref{kpv}), (\ref{ffeq}), (\ref{ffeq2}), and (\ref{inte}) the total energy 
of the system is 
\begin{align}
  E = \int \frac 1 2 [{\boldsymbol \nabla} \Phi({\bf r})]^2 \, d {\bf r} +&
\nonumber \\
  + \int \left[ \frac 1 2 (\Omega^2 + \omega^2) r^2 
  + a \Omega +  b \frac{\omega^2} {2 \Omega} \right] \Phi^2({\bf r}) \, d {\bf r} +&
  \nonumber \\
 + \int \frac 1 2 \frac {\Phi^4({\bf r})} {(1 - \zeta/2)^2} 
  \left[ \left( 1 - \frac 2 3 \zeta \right)
   \ln \frac {\Phi^2({\bf r})} {(1 - \zeta/2) \sqrt{e}} - \frac {\zeta} 9 \right] \, d {\bf r}.&
\nonumber \\
\label{ene}
\end{align}

Equation (\ref{ke4}), with the ansatz of Eq.\,(\ref{ansatz}), implies that 
\begin{equation}
  a(\zeta) = \frac {1 - \ln {\sqrt \zeta}} {1 - \zeta/2}.
\end{equation}
Also, Eqs.\,(\ref{ke5}) and (\ref{ansatz}) imply that
\begin{equation}
  b(\zeta) = \frac {\langle \rho_{c}^2 \rangle} {\ell_{\rm cell}^2} = \frac {1 - \zeta^2/3} {2 - \zeta}.
\end{equation}

We turn to the angular momentum, which is given by
\begin{equation}
  L = \int \Phi^2({\bf r}) \, f^2 \, ({\bf r} \times {\boldsymbol \nabla} \phi)_z \, d {\bf r}.
\label{angmom}
\end{equation}
Making similar approximations as before, it turns out that
\begin{equation}
  L = \pi \ell^2_{\rm cell} \sum_j \Phi^2({\bf R}_j) (\Omega R_j^2 + 1 ), 
\label{angmomm}
\end{equation}
which may again be converted into an integral,
\begin{equation}
   L = \int \Phi^2({\bf r}) (\Omega r^2 + 1) \, d{\bf r}.
\label{angmommm}
\end{equation}
Therefore, from Eqs.\,(\ref{ene}) and (\ref{angmommm}) it follows that the energy functional
$E - L \Omega - \mu N$ is 
\begin{align}
  {\cal E}(\Phi, \Phi^*) = E - L \Omega - \mu N = \int \frac 1 2 [{\boldsymbol \nabla} \Phi({\bf r})]^2 
  \, d {\bf r} +& \nonumber \\
 + \int \left[ \frac 1 2 (\omega^2 - \Omega^2) r^2 + (a - 1) \Omega  
 + b \frac {\omega^2} {2 \Omega} - \mu  \right] \Phi^2({\bf r}) \, d {\bf r} +& 
 \nonumber \\
 + \int \frac 1 2 \frac {\Phi^4({\bf r})} {(1 - \zeta/2)^2} 
  \left[ \left( 1 - \frac 2 3 \zeta \right)
   \ln \frac {\Phi^2({\bf r})} {(1 - \zeta/2) \sqrt{e}} - \frac {\zeta} 9 \right] 
   \, d {\bf r}.&
\nonumber \\
\label{enee}
\end{align}

It is useful to get some insight into the terms that appear in the above equation. 
The relevant dimensionless parameter is $N \omega$, which is $\approx R^2/a_0^2 \gg 1$, 
with $R$ being the radius of the droplet and $a_0 = 1/\sqrt{\omega}$ the oscillator 
length. (As we see below, in the Thomas-Fermi limit, $R^2 \sim N$). The three dominant 
terms in Eq.\,(\ref{enee}) are: The energy associated with the center-of-mass motion, 
$K_{\rm COM} = I \Omega^2/2$, where $I = (1/2) \int n_s({\bf r}) r^2 \, d {\bf r}$ 
is the moment of inertia of the droplet. The second dominant term is the energy due 
to the harmonic trapping potential, $V = I \omega^2/2$. Finally, the third dominant 
term is the $-L \Omega$ term, (i.e., the usual term that we use when we evaluate the 
energy in the rotating frame), where $L = I \Omega$. Obviously this term is equal to
$- I \Omega^2$. All these three terms are of order $N^2 \omega^2$ (here we assume that 
$\Omega \approx \omega$ and actually, for $\Omega = \omega$, $K_{\rm COM} + V - L \Omega$ 
vanishes, to leading order in $N \omega$). 

The energy due to the nonlinear term is of order $N$. Finally, the terms $(a - 1) \Omega$ 
and $b \omega^2/(2 \Omega)$ are of order $N \omega$. We should also mention that the 
kinetic energy that results from the spatial variations of $n_s({\bf r})$, $[{\boldsymbol 
\nabla} \Phi({\bf r})]^2$ is the smallest one and is of order $N/R^2$, i.e., of order unity
and is significant only at the edge of the droplet.

The (nonlinear) differential equation that follows from the energy functional of 
Eq.\,(\ref{enee}) is
\begin{align}
  \left( - \frac 1 2 {\boldsymbol \nabla}^2 + \frac 1 2 (\omega^2 - \Omega^2) r^2 
  + (a - 1) \Omega + b \frac {\omega^2} {2 \Omega} \right) \Phi({\bf r})+&
  \nonumber \\
  + \frac {\Phi^2({\bf r})} {(1 - \zeta/2)^2} \left[ \left(1 - \frac {2 \zeta} 3 \right)
  \ln \frac {\Phi^2({\bf r})} {(1 - \zeta/2) \sqrt{e}} + \right.&
  \nonumber \\
  + \left. 
  \frac 1 2 - \frac {4 \zeta} 9 \right] 
  \Phi({\bf r}) = \mu \Phi({\bf r}).&
\nonumber \\
\label{eqq}
\end{align}
From the solution of the above equation we get all the relevant parameters of the (rapidly
rotating) droplet. First of all, the value of $\zeta$ that minimizes the energy functional 
gives the fractional area of the vortex core in the unit cell. Also, the smoothed density 
distribution, $n_s({\bf r})$, is the solution of Eq.\,(\ref{eqq}). Furthermore, the angular 
momentum is given by Eq.\,(\ref{angmommm}), which is of order $N^2 \Omega \gg 1$, while the 
number of vortices $N_v \approx \Omega R^2$, i.e., $N_v \sim L/N \sim N \Omega \gg 1$.    

\section{Analytic results in the Thomas-Fermi limit, for $\Omega = \omega$}

In the Thomas-Fermi limit the first term in Eq.\,(\ref{eqq}) is negligible, and as a result 
this equation becomes an algebraic equation. For $\Omega = \omega$, this equation takes the 
even simpler form
\begin{align}
  \frac {\Phi^2({\bf r})} {(1 - \zeta/2)^2} \left[ \left(1 - \frac {2 \zeta} 3 \right)
  \ln \frac {\Phi^2({\bf r})} {(1 - \zeta/2) \sqrt{e}} + \right.&
  \nonumber \\
  + \left. \frac 1 2 - \frac {4 \zeta} 9 \right] = \mu_{\rm eff},&
\label{eqq2}
\end{align}
where $\mu_{\rm eff} = \mu - (a - 1) \omega - b \omega/2$. The solution of the above
equation gives the constant density of the droplet, while the normalization condition
determines its radius. 

\begin{figure} 
\includegraphics[width=\columnwidth]{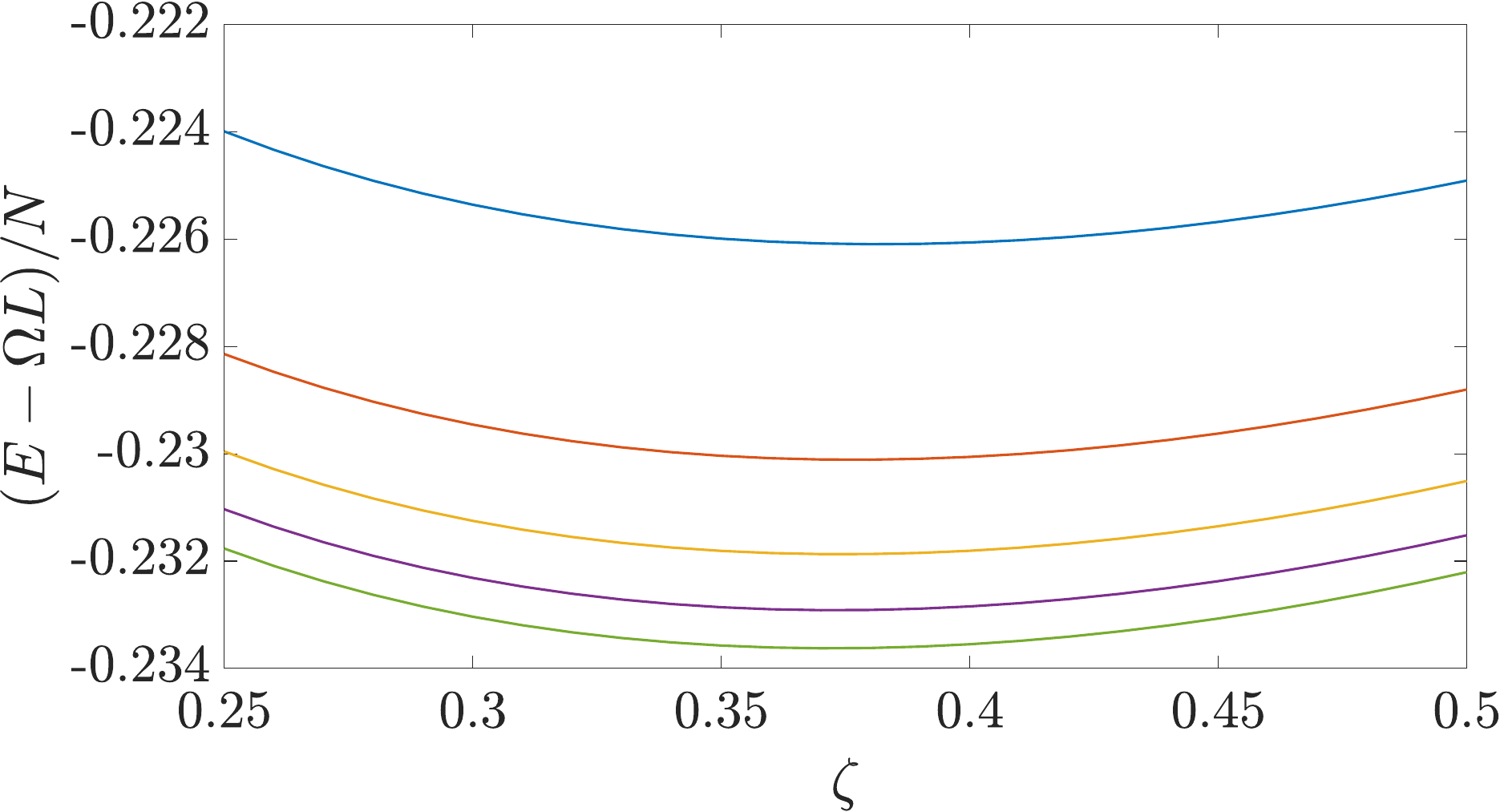}
\caption{The energy per particle in the rotating frame, $(E - L \Omega)/N$, as function 
of the variational parameter $\zeta$, for $\Omega = \omega = 0.05$ and $N = 1000, 2000, 
3000, 4000$, and 5000 from top to bottom. Here the energy is measured in units of $E_0$.}
\end{figure} 

We should recall at this point that for the non-rotating droplet, in the absence of any 
confining potential, and in the Thomas-Fermi limit, the energy functional is
\begin{align}
{\cal E}(\Phi, \Phi^*) = \int \frac 1 2 \Phi^4({\bf r}) 
\ln \frac {\Phi^2({\bf r})} {\sqrt e} \, d {\bf r}.
\label{eqqee}
\end{align}
The corresponding term of Eq.\,(\ref{enee}) reduces to the above expression when $\zeta = 0$.
The density of the non-rotating droplet results from minimizing the energy per particle 
that comes from Eq.\,(\ref{eqqee}), and is equal to $n_0 = 1/\sqrt{e}$. Imposing the 
normalization condition,
\begin{equation}
 R_0 = \left( \frac {N \sqrt{e}} {\pi} \right)^{1/2} \approx 0.72 \, {\sqrt N}.
\label{radiustf}
\end{equation}

In the present problem, minimization of the (interaction) energy per particle implies
that the smoothed (flat-top) density of the droplet $n_s(\zeta)$ is,
\begin{equation}
   n_s(\zeta) = n_0 (1 - \zeta/2) e^{\zeta/(9 - 6 \zeta)},
   \label{RRR}
\end{equation} 
which is a decreasing function of $\zeta$, for $0 < \zeta < 1$, as expected.
Also, the corresponding radius $R(\zeta)$ is,
\begin{equation}
   R(\zeta) = R_0 \left(  
   \frac {e^{-\zeta/(9 - 6 \zeta)}} {1 - \zeta/2} \right)^{1/2},
   \label{RR}
\end{equation} 
which is an increasing function of $\zeta$, in the same interval.
In Fig.\,1 we show $(E - L \Omega)/N$ as function of $\zeta$, for $\Omega = \omega$, which 
we find numerically. Here $N = 1000$, 2000, 3000, 4000, and 5000 from top to bottom. In this
plot we see that the value of $\zeta$ which minimizes the energy in the rotating frame 
approaches the value $\zeta_0 \approx 0.37$. For this value, 
\begin{equation}
  R(\zeta_0) \approx 1.08 \, R_0 \approx 0.78 \sqrt{N},
  \label{r00}
\end{equation}
and 
\begin{equation}
 n_s(\zeta_0) \approx 0.86 \, n_0. 
 \label{n00}
\end{equation}
Both the increase in $R(\zeta)$ (by roughly 8\%) and the decrease in $n_s(\zeta)$ (by roughly
14\%) as compared to the non-rotating droplet are due to the presence of the vortices. We stress 
that in the derivation of the above results we have kept only the energy due to the nonlinear 
term, which is of order $N$. Among the neglected terms two are the most important, i.e., the 
ones associated with $a$ and $b$, which are of order $N \omega$. These are also of order $N$, 
but much smaller due to the assumption $R^2/a_0^2 \gg 1$, which we discussed in the previous
section. It is needless to say that in the numerical results none of the above terms is 
neglected.

Finally, regarding the size of each vortex core $\xi$,
\begin{equation}
  \frac {\xi} {R(\zeta_0)} \sim \frac 1 {\sqrt {N \omega}}.  
\end{equation}
As expected, this ratio is $\ll 1$, since $N \omega \gg 1$.

\begin{figure}[t]
\includegraphics[width=\columnwidth]{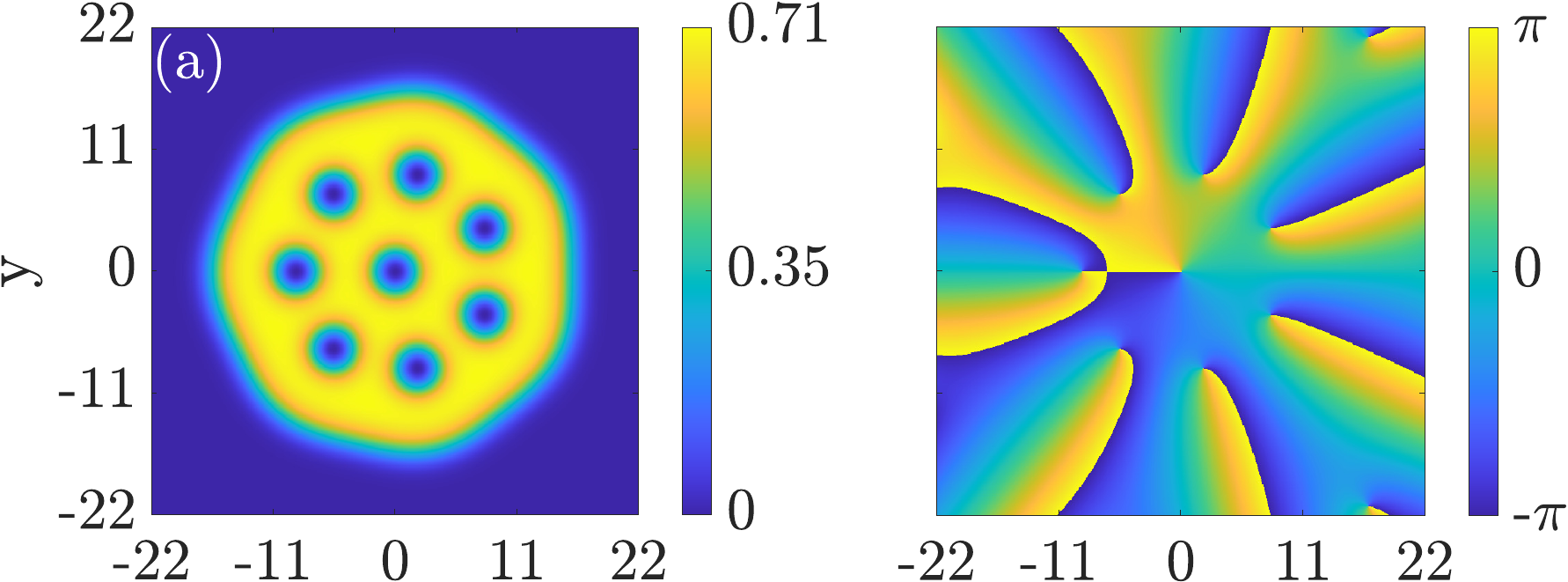}\\
\vspace{0.5\baselineskip}
\includegraphics[width=\columnwidth]{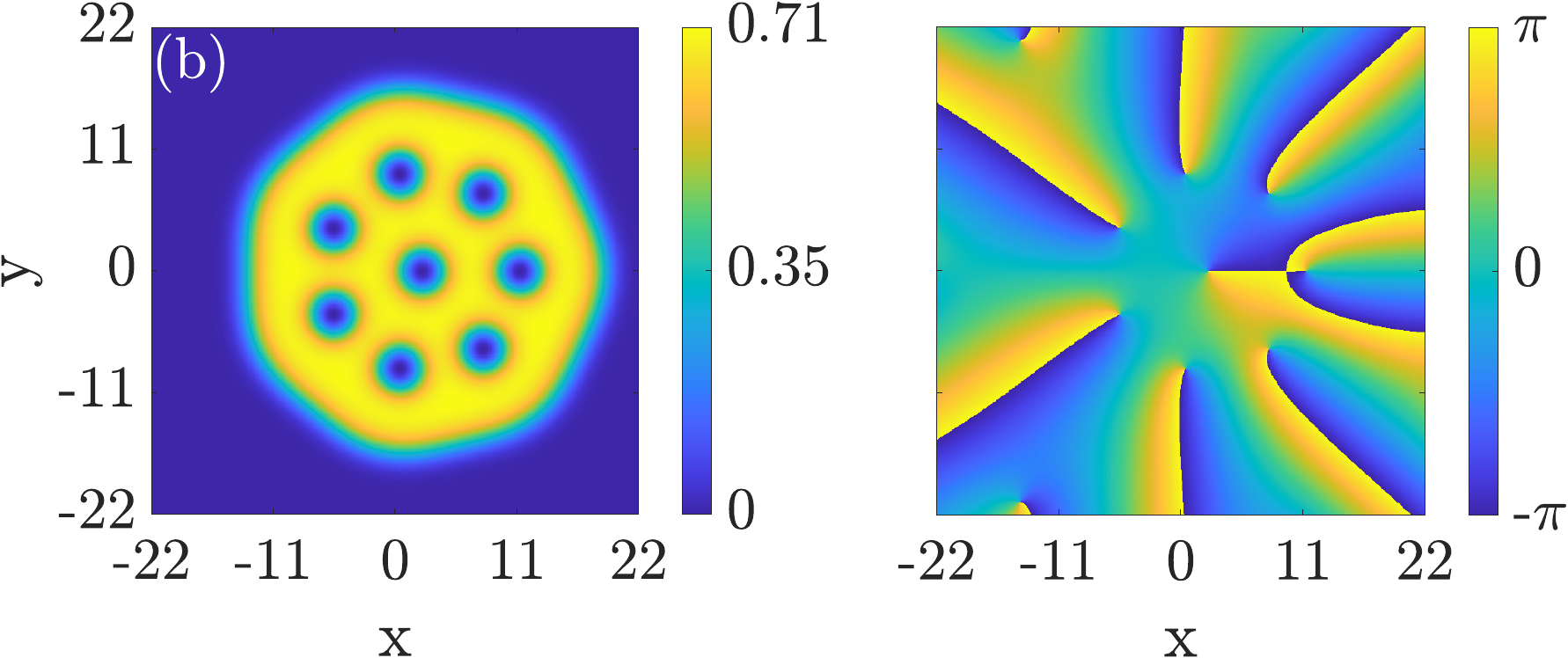}
\caption{The density (left column) and the phase (right column) of the droplet order parameter for $N = 500$, $\omega = 0.05$, and (a) $L/N = 6.1$, and (b) $L/N = 6.4$. Here the density is measured in units of $\Psi_0^2$ and the length in units of $x_0$.}
\end{figure} 

\section{Connection between center-of-mass and vortex excitation}

Let us start with the first main question of this study, i.e., how the angular momentum is
distributed between vortex and center-of-mass excitation. To answer this question we will
rely on the combination of two approaches. The first one is the minimization of the energy 
functional of Eq.\,(\ref{funncc}), using the damped second-order-in-fictitious-time method \cite{GO}, which is suitable in the limit of relatively 
small droplets. As we have seen in Ref.\,\cite{NKO}, for a fixed $\omega = 0.05$, the 
transition from pure center-of-mass excitation to vortex excitation takes place for the
critical value of $N$ between 98.6 and 98.7. Up to this value of $N$, the droplet carries
its angular momentum via center-of-mass excitation (only) for any value of $\ell = L/N \ge 0$.
Denoting as $\ell_0$ the critical value of $L/N$ above which we have the transition from
center-of-mass excitation to vortex excitation, we thus find that for any value of $N$
up to $\approx 98.6$, $\ell_0 = 0$. 

\begin{figure}[t]
\includegraphics[width=\columnwidth]{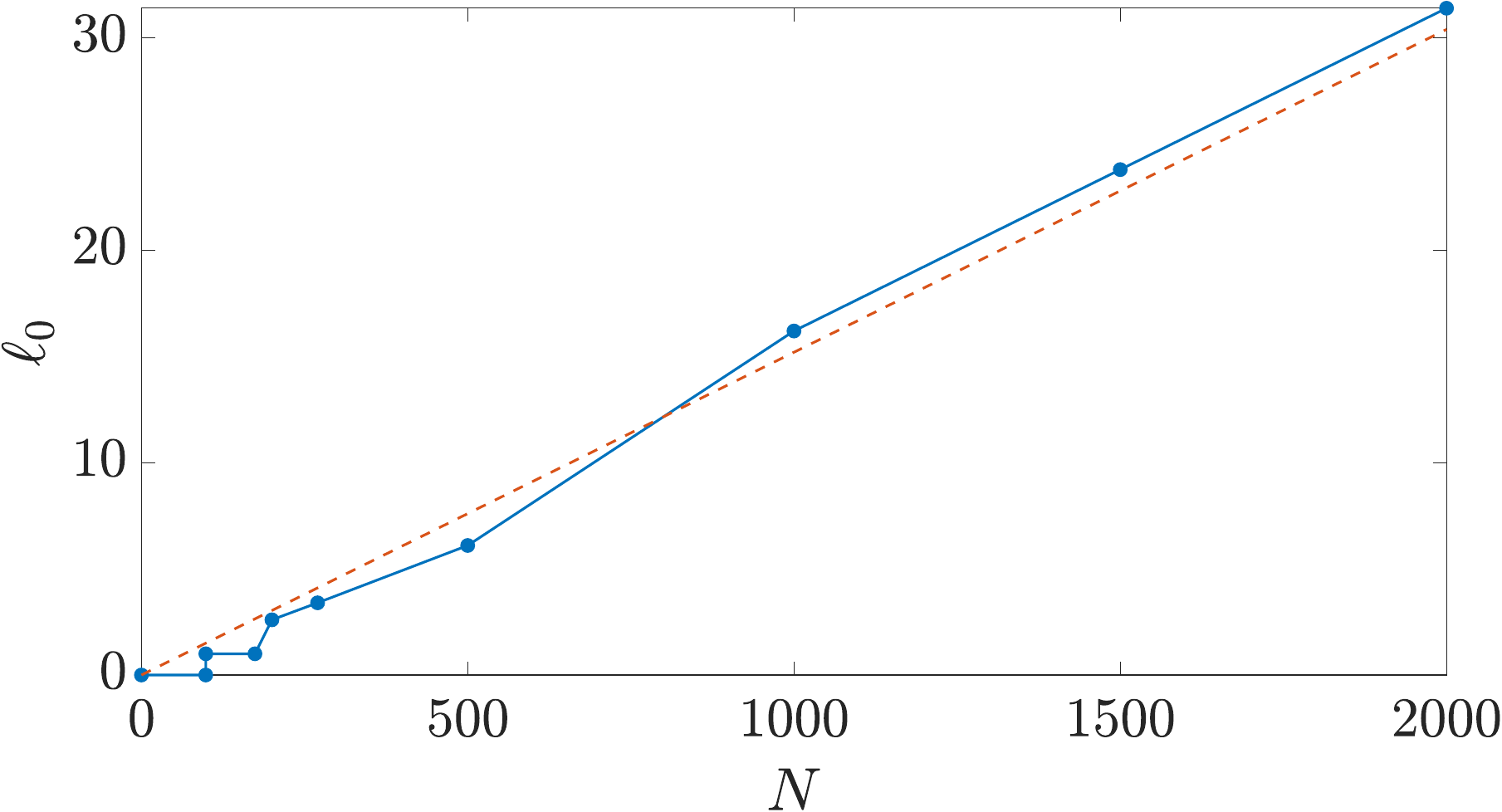}
\caption{The critical value of the angular momentum per particle (in units of $\hbar$) for the transition between vortex
and center-of-mass excitation as function of $N$. Alternatively one may view this as the phase 
diagram which involves vortex excitation in the lower part and center-of-mass excitation in the 
upper part. The dashed straight line is the approximate result of Eq.\,(\ref{approxl}) and the 
solid curve with data points is the full numerical result.}
\end{figure} 

In the same study we have found that for $N = 100$, the value of $\ell_0$ is equal to 
unity (in which case the droplet already has a singly-quantized vortex state at its 
center). For $N = 200$, the value of $\ell_0$ becomes $\approx 2.6$, while for $N = 270$, 
$\ell_0 \approx 3.4$. Finally, within the present study, we have managed to extract the
value of $\ell_0$ for $N = 500$, which is $\approx 6.1$. For this value of the angular momentum, which corresponds to $\Omega = \omega = 0.05$, the droplet exists in a state which carries eight singly-quantized vortices, as shown in Fig.\,2(a). For higher values of the angular momentum, e.g. $L/N = 6.4$, the droplet turns to center-of-mass excitation of the vortex carrying state, as shown in Fig.\,2(b). We stress that for $L/N \geq 6.1$ the dispersion relation (i.e., the energy as a function of $L$) becomes linear, with a slope equal to $\omega = 0.05$, as we have seen in Ref.\,\cite{NKO}. 

For even larger values of $N$ this calculation becomes increasingly difficult. We stress
that fixing the angular momentum (as compared with fixing $\Omega$) introduces an extra 
constraint, which makes the numerical calculation more demanding. Thus, in order to 
evaluate $\ell_0$ for higher values of $N$, we rely on the Wigner-Seitz approximation, 
which was described in Sec.\,IIB. Since within this method we minimize the energy in the 
rotating frame at fixed $\Omega$, therefore $\Omega = {\partial E(L)}/{\partial L}$.
In other words, for some given $\Omega$ that we choose, we get the slope of the dispersion 
relation for a value of $\ell$, which is precisely $\ell_0$. However, we know that we have 
the transition to center-of-mass excitation when $\Omega$ becomes equal to $\omega$. 
Then, all that remains to be done is to examine the dependence of $\ell_0$ on $N$. 

\begin{figure}[!ht]
\includegraphics[width=\columnwidth]{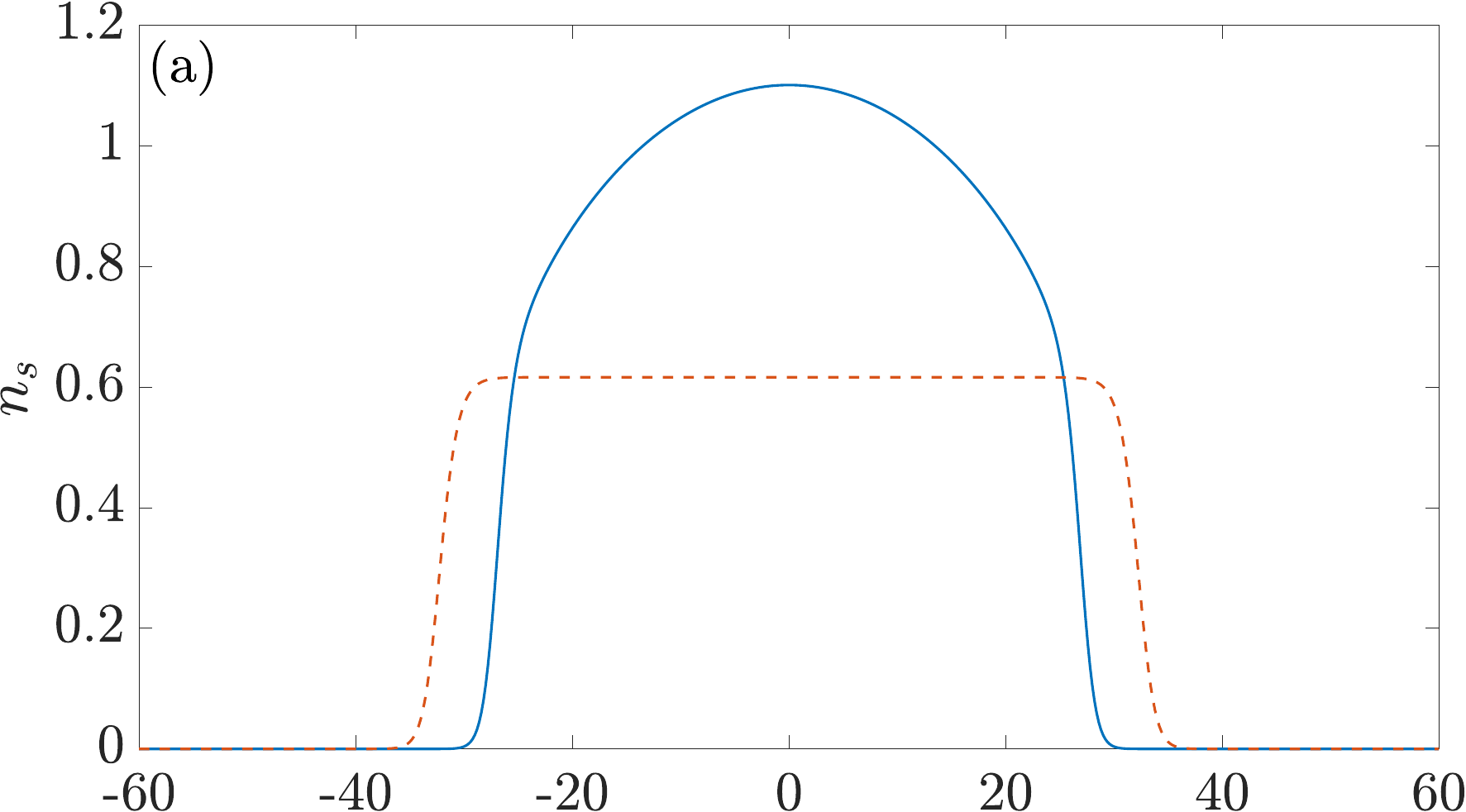}\\
\vspace{0.5\baselineskip}
\includegraphics[width=\columnwidth]{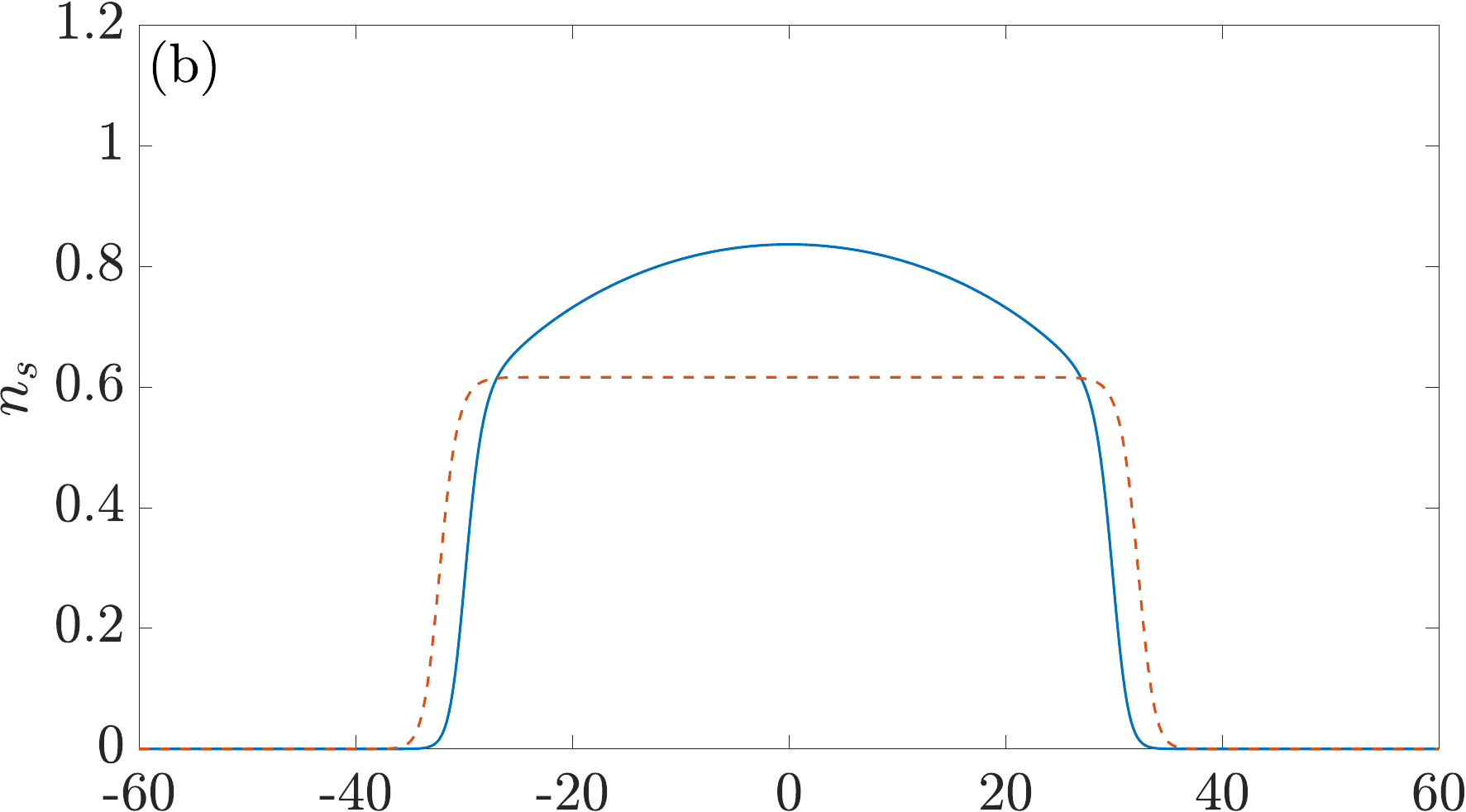}\\
\vspace{0.5\baselineskip}
\includegraphics[width=\columnwidth]{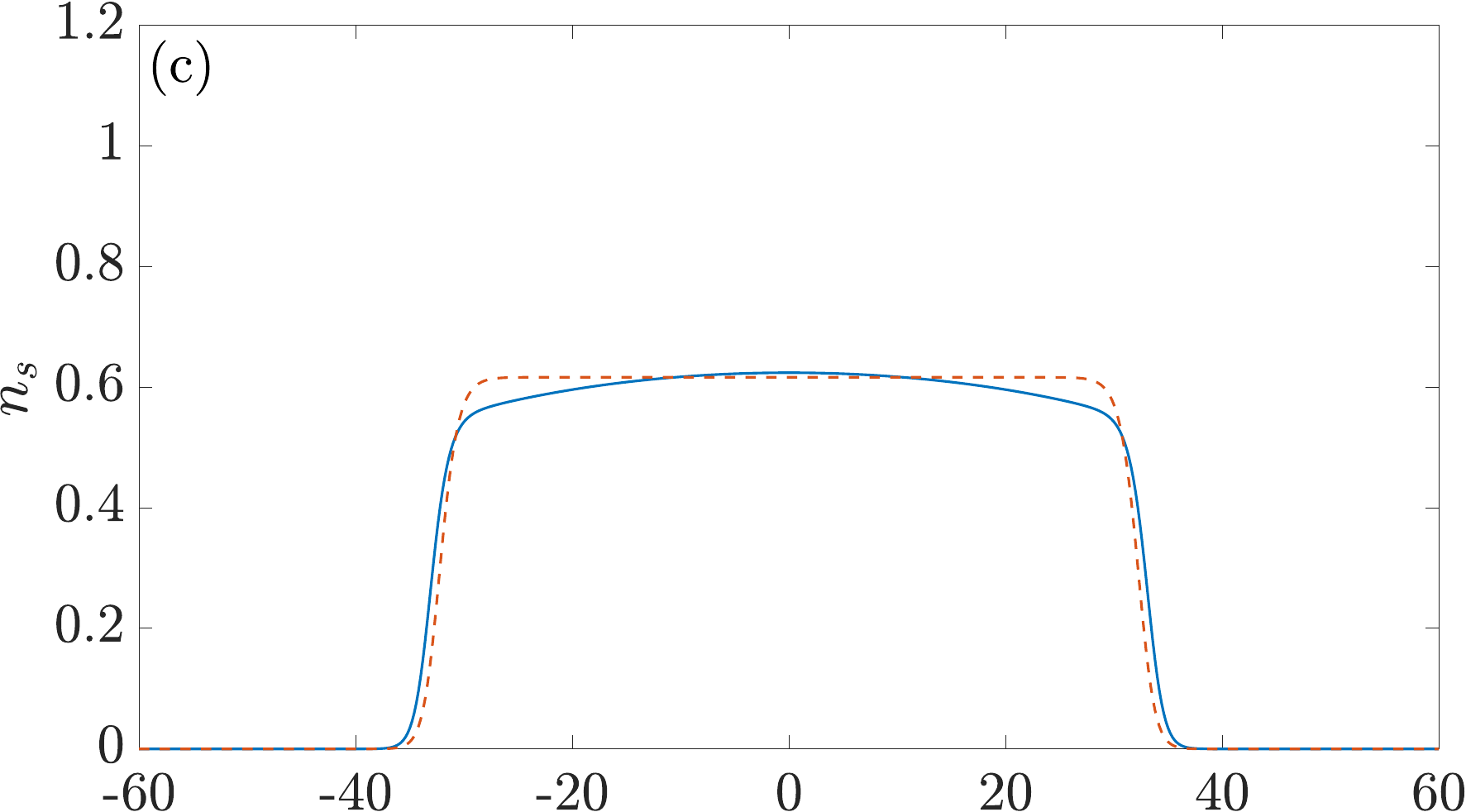}\\
\vspace{0.5\baselineskip}
\includegraphics[width=\columnwidth]{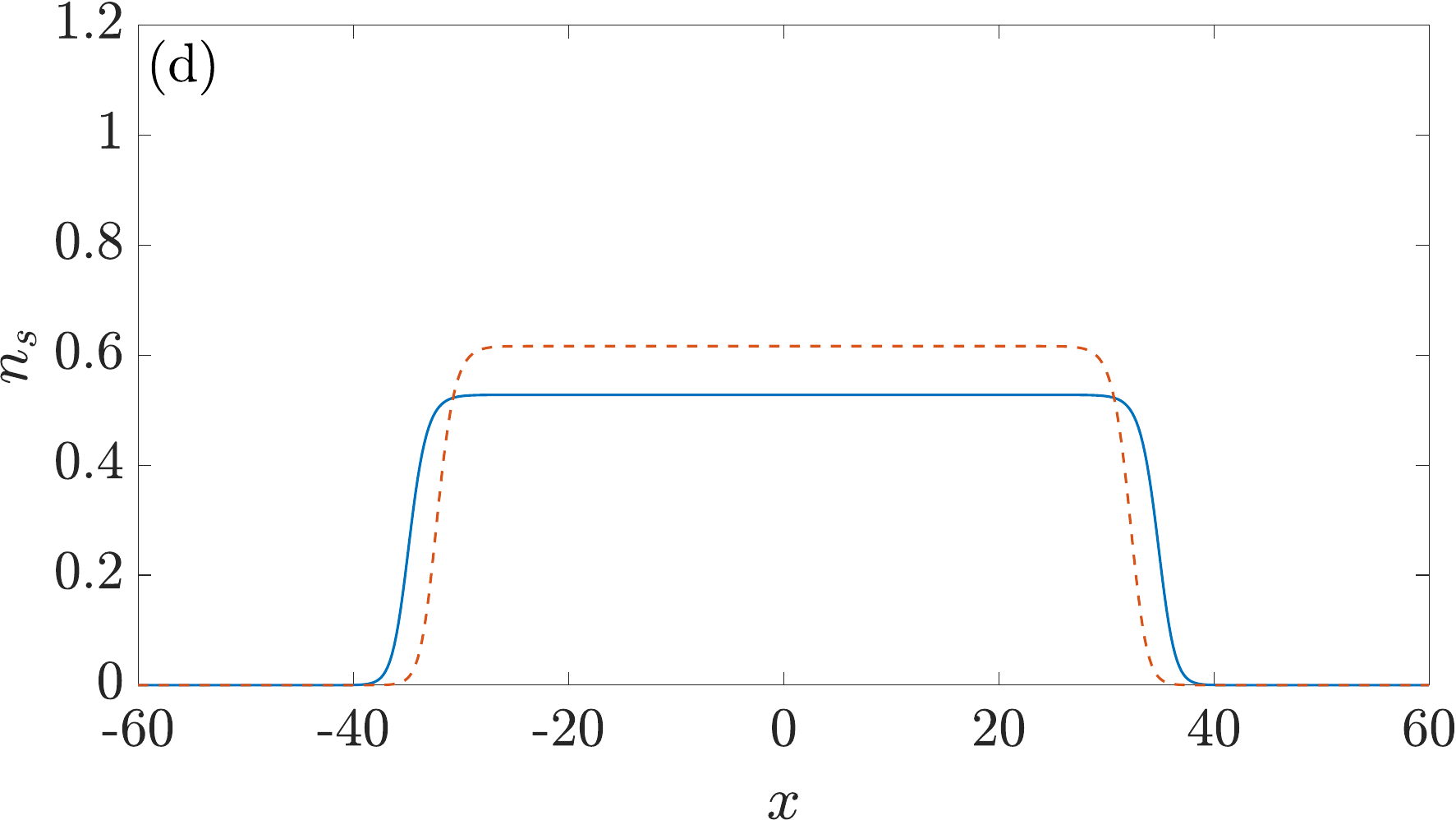} 
\caption{The smoothed density distribution, $n_s({\bf r}) = \Phi^2({\bf r})$, of a quantum droplet, for (a) $\Omega = 0.035$, (b) $\Omega = 0.045$, (c) $\Omega = 0.049$, and (d) $\Omega = 0.05$. The dashed curve shows the density of the non-rotating, unconfined droplet. In all the
plots $N=2000$. Here the density is measured in units of $\Psi_0^2$ and the length in units of $x_0$.}
\end{figure} 

\begin{figure} 
\begin{center} 
\includegraphics[width=\columnwidth]{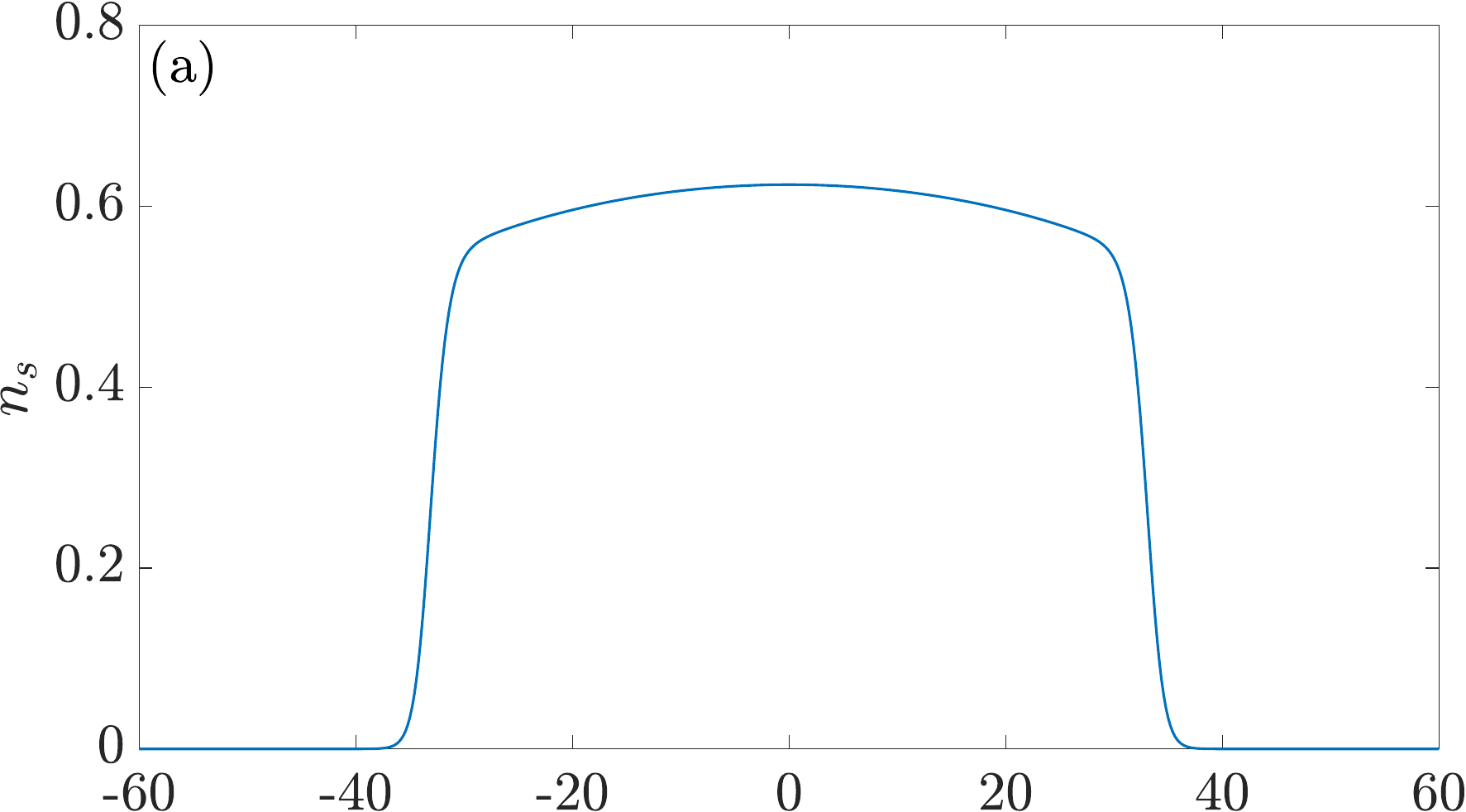}\\
\vspace{0.5\baselineskip}
\includegraphics[width=\columnwidth]{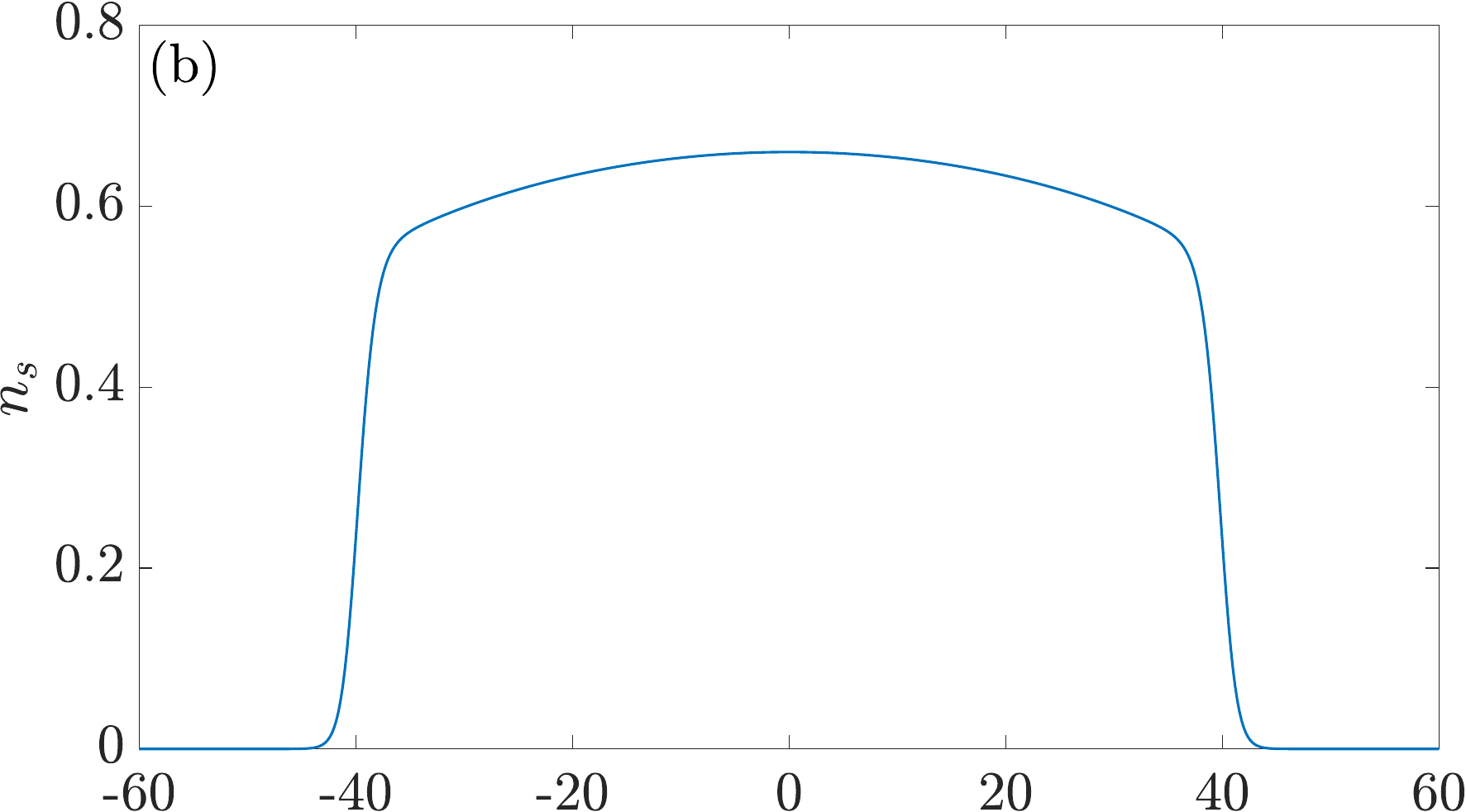}\\
\vspace{0.5\baselineskip}
\includegraphics[width=\columnwidth]{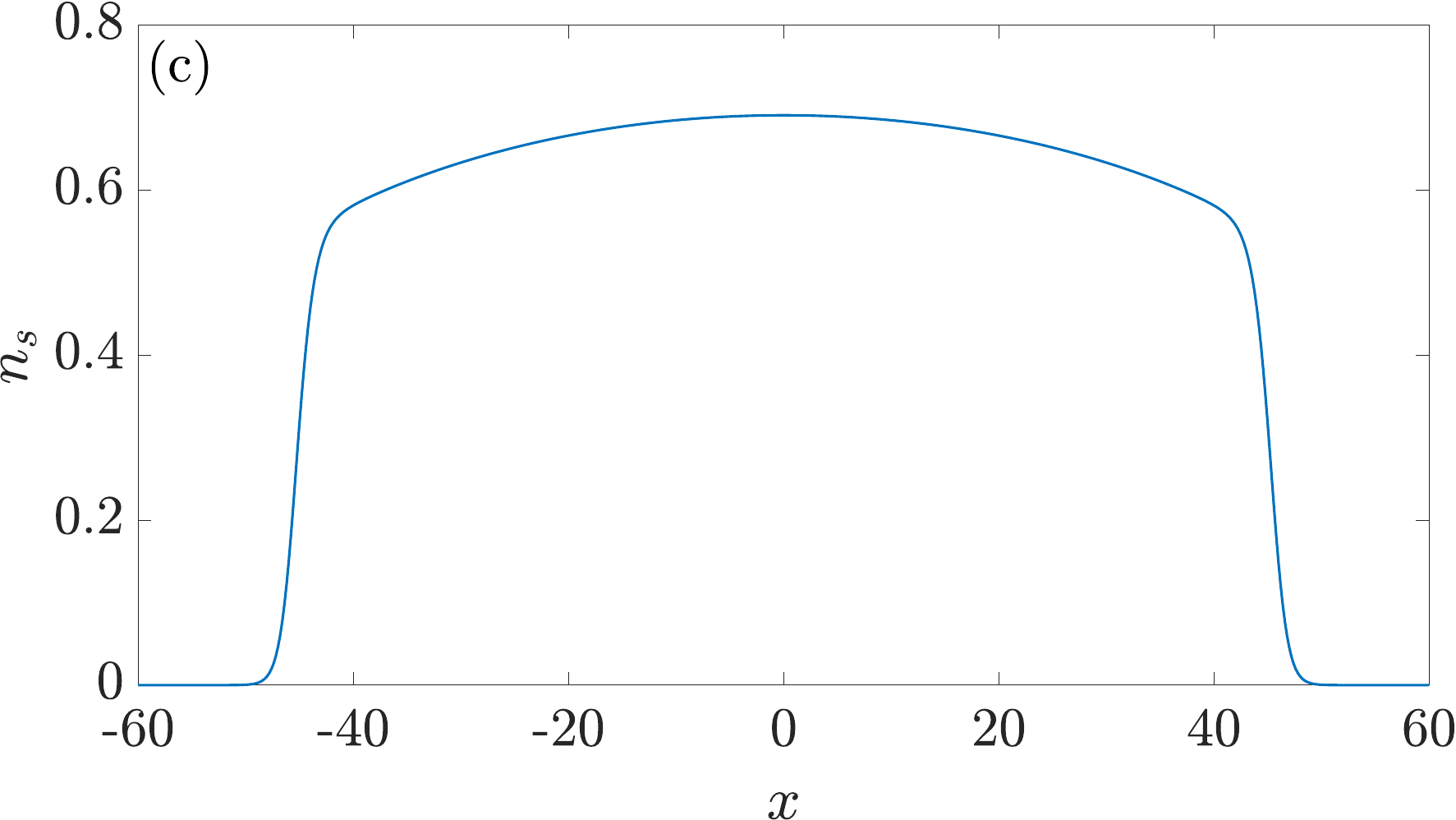}
\end{center} 
\caption{The smoothed density distribution, $n_s({\bf r}) = \Phi^2({\bf r})$, 
of a quantum droplet, for (a) $N=2000$, (b) $N=3000$, and (c) $N=4000$.
In all the plots $\Omega = 0.049$. Here the density is measured in units of $\Psi_0^2$ and the length in units of $x_0$.}
\end{figure} 

Actually, from Eqs.\,(\ref{angmommm}), (\ref{radiustf}), and (\ref{r00}) we find that
\begin{equation}
  \ell_0 = \frac 1 2 \omega R^2(\zeta_0) \approx 0.0152 \, N,
  \label{approxl}
\end{equation}
i.e., there is an approximately linear dependence of $\ell_0$ on $N$. The number of vortices $N_v$ may also be
evaluated using Feynman's relation along with Eq.\,(\ref{r00}),
\begin{equation}
  N_v = n_v \pi R^2(\zeta_0) = \omega R^2(\zeta_0) = 2 \ell_0 \approx 0.0304 \, N. 
\end{equation}
Figure 3 shows the $\ell_0$ values for a ``small" droplet ($N = 98.6, 98.7, 174, 200, 270$ and 500), which  
result from minimizing the energy functional of Eq.\,(\ref{funncc}) \cite{NKO}. The rest of 
the points ($N = 1000, 1500$, and 2000) were derived from the Wigner-Seitz approximation. Their $\ell_0$ values are $16.2$, $23.8$ and $31.4$, respectively. For $0 \leq N \leq 98.6$ we see that $\ell_0 = 0$, as for that range of $N$ values the droplet carries its angular momentum only via center-of-mass excitation. Following that, the curve exhibits another horizontal region, for $98.7 \leq N \leq 174$, where $\ell_0 = 1$. For this range of $N$ values, the droplet accommodates exactly one singly-quantized vortex before turning to center-of-mass excitation. The curve then develops a more detailed structure, as the droplet accommodates two, or more, vortices. Finally, for large $N$ values the curve turns linear. We stress here that the slope of the curve in that region, as calculated through our numerical results, is in excellent agreement with the semi-analytic value given by Eq.\,(\ref{approxl}).

\begin{figure} 
\begin{center} 
\includegraphics[width=\columnwidth]{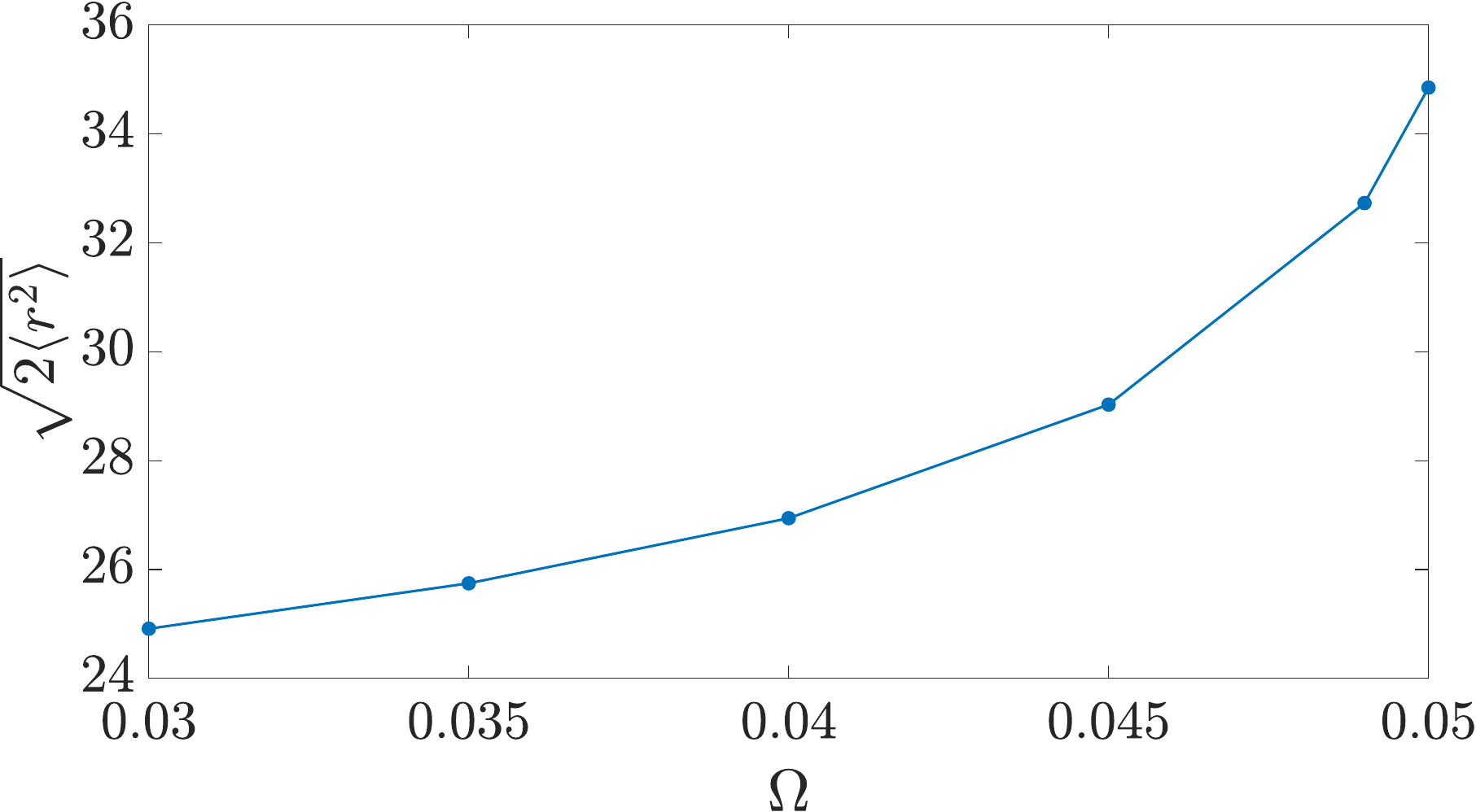}
\end{center} 
\caption{The expectation value $\sqrt{2 \langle r^2 \rangle}$ (in units of $x_0$) as function of $\Omega$ (in units of $\omega_0$), 
for $\Omega = 0.03, 0.035, 0.04, 0.045, 0.049$ and $0.05$. Here $N=2000$.}
\end{figure}

\begin{figure} 
\begin{center} 
\includegraphics[width=\columnwidth]{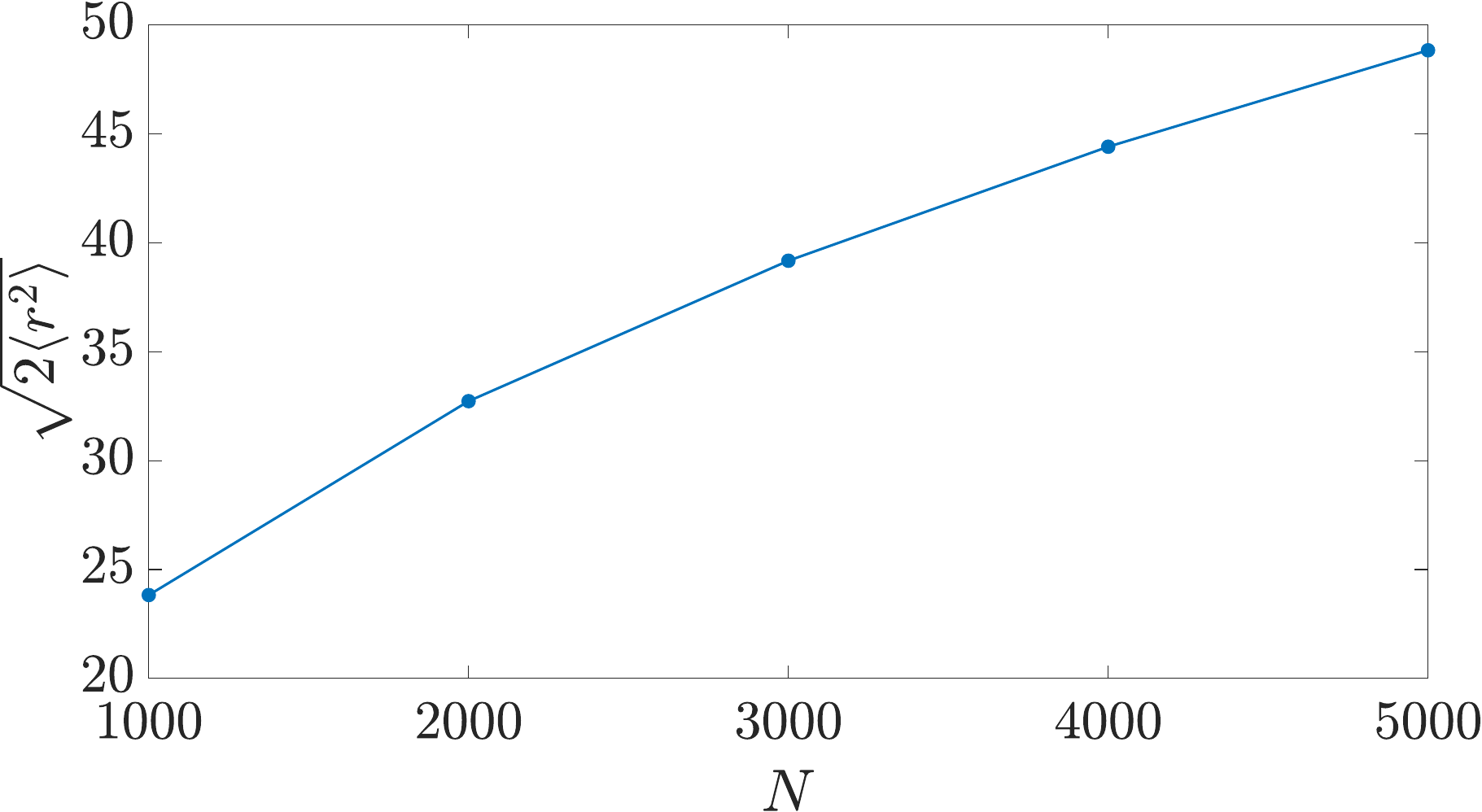}
\end{center} 
\caption{The expectation value $\sqrt{2 \langle r^2 \rangle}$ (in units of $x_0$) as function of $N$, 
for $N=1000, 2000, 3000, 4000$, and $5000$. Here $\Omega = 0.049$.}
\end{figure} 

\section{Properties of the vortex lattice}

In the previous section we focused on the case where $\Omega = \omega$. Now we turn 
to the second main question of this study, namely the properties of the vortex 
lattice as $\Omega$, or as $N$ are varied. All the results presented below come from 
the Wigner-Seitz approximation. 

The first question that we examine is the smoothed density distribution, $n_s({\bf r}) 
= \Phi^2({\bf r})$, of the droplet. We derive this from the minimization of the energy
functional of Eq.\,(\ref{enee}) for two cases. The first one is to fix $N$ to 2000 and 
vary $\Omega$ from 0.03 up to 0.05, which is the value of $\omega$. In all these cases 
we minimize the energy with respect to our variational parameter, which is $\zeta$. As 
seen in Fig.\,4, there is a gradual transition of the density to a ``flat top" 
distribution as $\Omega$ increases and reaches $\omega$. Actually, the density 
distribution for $\Omega = \omega$ is very much like the one of the non-rotating, unconfined 
droplet, with the only difference being that the droplet has expanded radially 
[see Eq.\,(\ref{r00})] and has a lower density [see Eq.\,(\ref{n00})].  

In the second case we fix $\Omega$ and vary $N$. Here we choose $\Omega = 0.049$ 
to be very close to $\omega = 0.05$ and we choose the values $N = 2000, 3000$ 
and 4000. Figure 5 shows the result of this calculation. In this case we observe that 
both the width of the droplet, as well as its height increase. 

In Fig.\,6 we plot the expectation value $\sqrt{2 \langle r^2 \rangle}$, which,
for a flat-top distribution gives the radius of the droplet as function of $\Omega$, 
for fixed $N = 2000$. As $\Omega$ approaches $\omega$ the effective potential 
$(\omega^2 - \Omega^2) r^2/2$ softens and as a result the droplet expands, reaching 
a maximum value when $\Omega = \omega$. Beyond this value of $\Omega = \omega$ the 
droplet undergoes center-of-mass excitation -- see Fig.\,3. In Fig.\,7 we plot the 
same quantity, $\sqrt{2 \langle r^2 \rangle}$, as function of $N$, for fixed $\Omega = 0.049$. Again, this is an increasing function of $N$, 
as expected.

In Figs.\,8 and 9 we plot the value of $\zeta = (\pi \xi^2)/(\pi \ell^2_{\rm cell})$
that minimizes the energy. Obviously $\zeta$ is the fractional area of the vortex size
over the cell size. In Fig.\,8 we plot $\zeta$ as function of $\Omega$ for $N = 2000$
and $\Omega$ from 0.03 up to 0.05, which is the value of $\omega$. We observe that 
$\zeta(\Omega)$ is an increasing function. Finally, Fig.\,9 shows $\zeta$ as function 
of $N$, for a fixed value of $\Omega = 0.049$. Here, $\zeta(N)$ decreases with increasing $N$, albeit relatively slowly.

\begin{figure}[t]  
\includegraphics[width=\columnwidth]{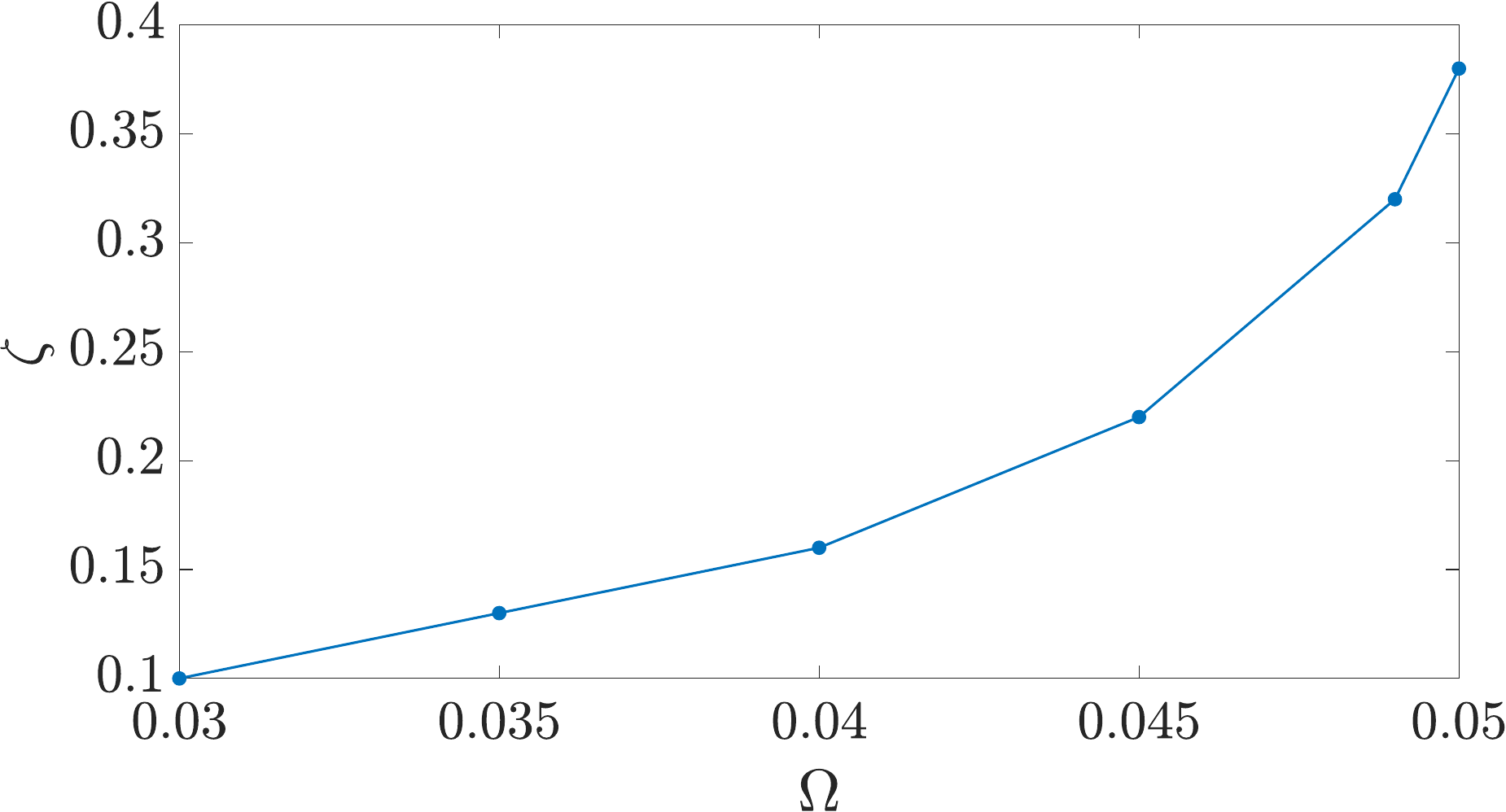} 
\caption{The fractional area $\zeta(\Omega)$ of each vortex size over the cell size 
as function of $\Omega$ (in units of $\omega_0$), for $\Omega = 0.03, 0.035, 0.04, 0.045$, $0.049$ 
and $0.05$. Here $N=2000$.}
\end{figure} 

\begin{figure}
\includegraphics[width=\columnwidth]{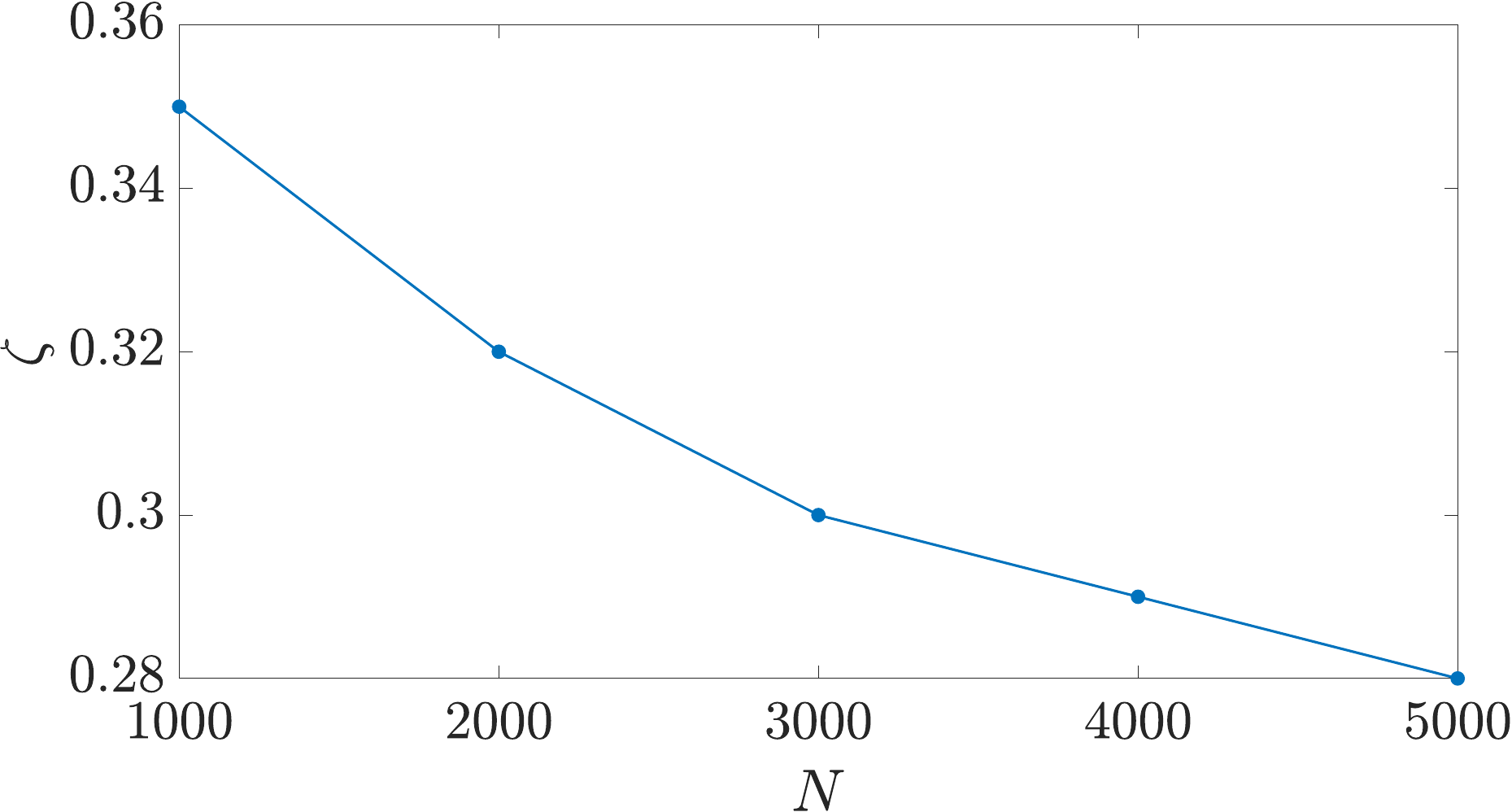}
\caption{The fractional area $\zeta(N)$ of each vortex size over the cell size as 
function of $N$, for $N=1000, 2000, 3000, 4000$, and $5000$. Here $\Omega = 0.049$.}
\end{figure} 

From Eq.\,(\ref{angmommm}) we plot in Fig.\,10 the angular momentum per particle 
$L(\Omega)/N$ as function of $\Omega$, considering $N = 2000$. This is an increasing 
function, as expected. Finally, in Fig.\,11 we present the total density (rather than the smoothed density) of the droplet order parameter, $n({\bf r})=\Psi^2({\bf r}) = \Phi^2({\bf r}) \cdot f^2({\bf r})$, in a region around the origin, for $\Omega=0.03$ and $0.05$. We observe that as $\Omega$ increases, the inter-vortex spacing $2\ell_{\rm cell}$ decreases, and the fractional area $\zeta$ of each vortex core over the cell increases. Here, we have considered the value $N=2000$, however we stress that this picture is representative of other (large) $N$ values as well, as the inter-vortex spacing  does not depend on $N$, and $\zeta$ depends relatively weakly on $N$.

\begin{figure} 
\includegraphics[width=\columnwidth]{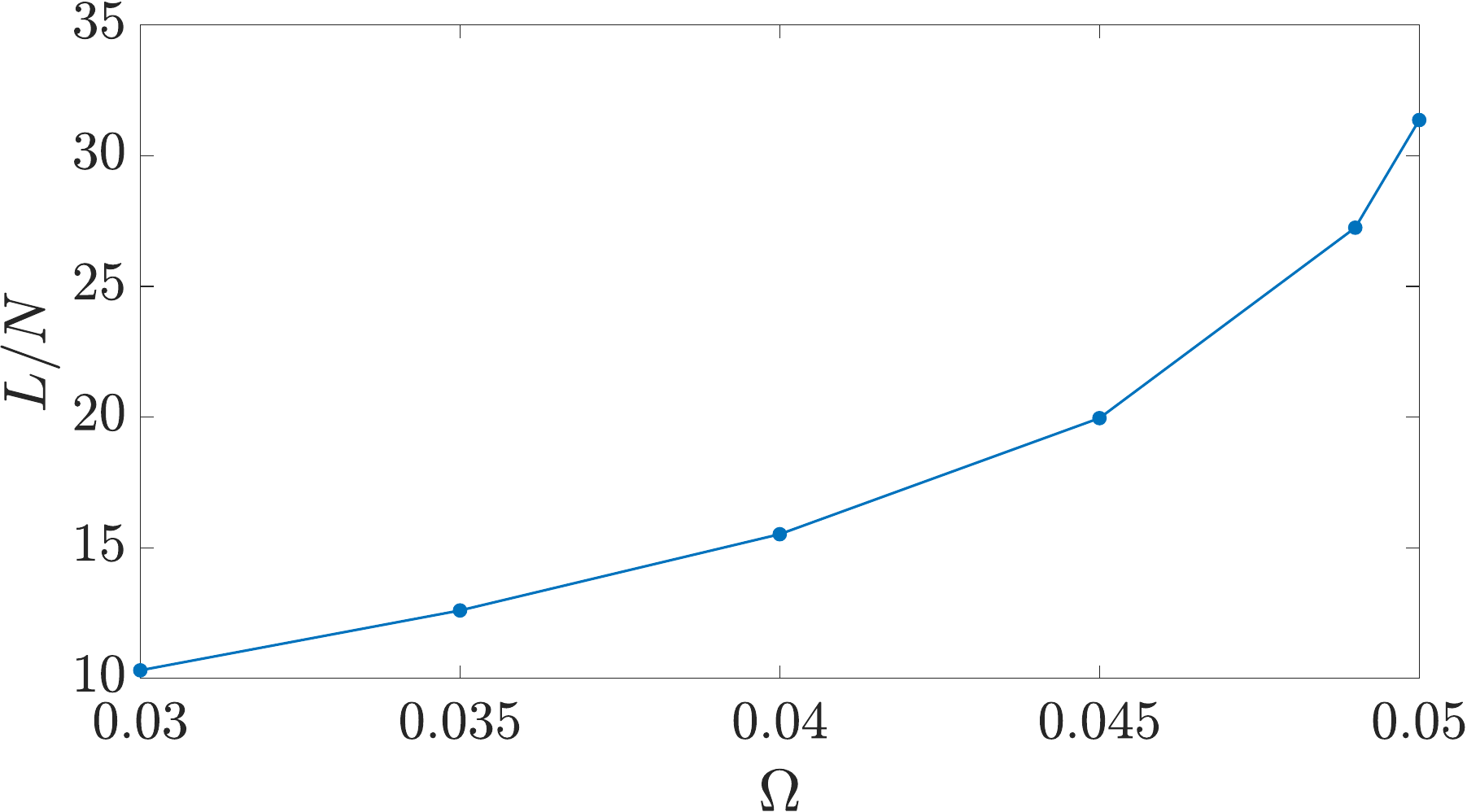}
\caption{The angular momentum per particle $L(\Omega)/N$ (in units of $\hbar$) as function of $\Omega$ (in units of $\omega_0$), 
for $\Omega = 0.03, 0.035, 0.04, 0.045$, $0.049$ and $0.05$. Here $N=2000$.}
\end{figure} 

\begin{figure} 
\includegraphics[width=\columnwidth]{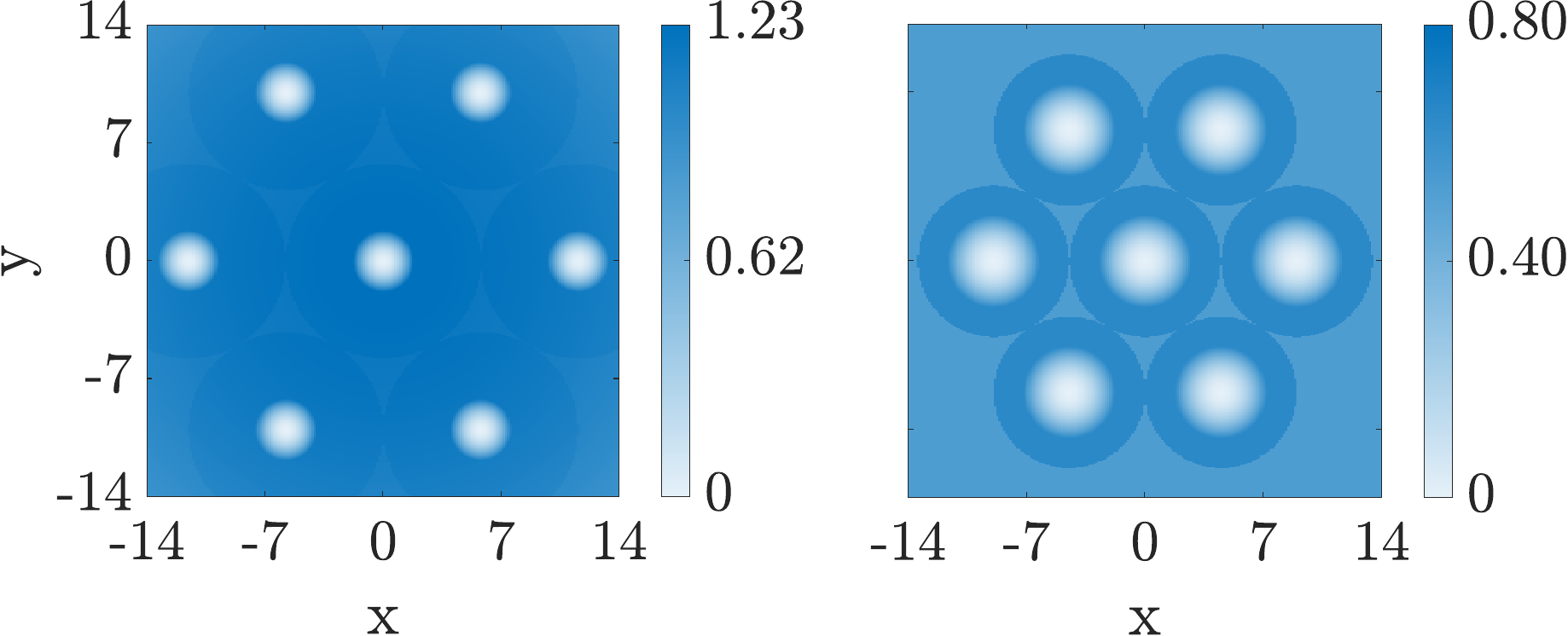}
\caption{The density of the droplet order parameter around the origin for $N = 2000$, and $\Omega = 0.03$ (left), and $\Omega = 0.05$ (right). Here the density is measured in units of $\Psi_0^2$ and the length in units of $x_0$.}
\end{figure} 

\section{Discussion of the results and summary}

In this study we considered the problem of a rapidly-rotating quantum droplet, which 
is confined in a harmonic potential, in purely two spatial dimensions. The harmonic 
potential we consider has two unique properties. The first one is the fact that in a 
harmonic potential the center-of-mass coordinate separates from the relative coordinates.
The second is that as the rotational frequency of the trap approaches the trap frequency 
the centrifugal potential cancels exactly the confining potential. 

In this study we combined two approaches, namely a full numerical minimization of the 
corresponding energy functional, as well as a Wigner-Seitz approximation for the case 
where there is a vortex lattice, following Refs.\,\cite{GB1, GB2, GB3, GB4, GB5, GB6}. 
We assumed that there is a smoothed, slowly-varying, density distribution and, on 
top of that, a rapidly-varying density distribution due to the presence of the vortices.
We also assumed that each vortex occupies a cylindrical cell, and we treated the size of 
the vortex core variationally. Since we considered this to be spatially independent, we 
expect that our result become more accurate when $\Omega$ approaches $\omega$, in which 
case the droplet has a flat-top density distribution. 
 
In one of the main results of our study, we managed to develop an equation for the smoothed 
density distribution. This equation resembles the corresponding one of the non-rotating 
droplet. The approach that we use not only allows us to deal with the asymptotic limit of 
a very large vortex lattice, but also we manage to derive some analytic results, which 
although are not exact, give insight into the problem.
  
The fact that the droplet is self-bound makes this problem very different as compared to 
the corresponding problem of a repulsive contact potential. In the problem with contact 
interactions -- which has been studied extensively in the past -- as $\Omega$ approaches 
$\omega$, the cloud expands and a vortex lattice forms. Eventually, the system enters a 
highly-correlated regime, where the lattice ``melts" and the many-body state develops 
correlations beyond the mean-field, product state \cite{rev1, rev2, rev3, rev4, rev5}. 

In the present problem the nonlinear term is partly attractive and partly repulsive. 
According to our results, as $\Omega$ approaches $\omega$ there may, or there may not be 
a vortex lattice, depending on the droplet atom number and the angular momentum, as shown in
Fig.\,3.  For a ``small" droplet there is a transition to center-of-mass excitation, with 
none, or few vortices \cite{NKO}. For a ``large" droplet we have the formation of a vortex 
lattice, which, however, never ``melts", even when $\Omega$ becomes equal to $\omega$. 
Actually, nothing really important happens in the droplet in this case, apart from the fact 
that the droplet has expanded and the density has dropped (compared with the non-rotating 
droplet), due to the presence of the vortices, as seen in Figs.\,4 up to 10. Regarding the 
area of each vortex core, compared with the size of each cell, this also increases with 
increasing $\Omega$. On the other hand, for fixed $\Omega$ we have seen a slow decrease of this ratio, as $N$ increases. 

In addition, when $\Omega$ becomes equal to $\omega$, the confining potential is identically 
canceled by the centrifugal potential. In this case and in the Thomas-Fermi limit of a large
droplet the physics is determined solely by the attractive term which makes it self-bound 
(this is the main difference with the case of repulsive contact interactions). In this respect, 
there is a universal behaviour of this problem, for $\Omega = \omega$ and for a ``large" droplet. 
More specifically, the droplet has a flat-top shape (apart from the nodes in the density, due to 
the presence of the vortices). 

Finally, when $\Omega$ exceeds $\omega$ -- even by an infinitesimal amount -- the energy of 
the droplet is not bounded any more. The same happens also for contact interactions. However, 
a droplet always turns to center-of-mass excitation, escaping to infinity, preserving its shape. 
This is due to the fact that the droplet is self-bound. In the problem of contact repulsive 
interactions, on the other hand, the atoms fly apart. 

As a final remark, we stress that the two approaches that we have followed, namely 
minimization of the energy fixing the angular momentum, or fixing the angular velocity 
of the trap are intimately connected. If the angular momentum is fixed, one may easily 
derive the results for fixed $\Omega$, however the reverse is not possible (at least 
in a direct way). Furthermore, the two approaches correspond to different experimental 
situations. More specifically, if the angular momentum is fixed, one would be able to 
observe the vortex-carrying droplet executing center-of-mass motion, without escaping 
to infinity and the reason is the constraint of a fixed angular momentum. On the other 
hand, if one works with a fixed $\Omega$, as soon as $\Omega$ exceeds $\omega$ -- even 
by an infinitesimal amount -- the vortex-carrying droplet would escape to infinity. 
Consequently, this would never be a stationary state (in the rotating frame), and the 
only chance to observe it experimentally would correspond to some ``transient" state.

From the above discussion it is clear that there is a whole collection of results
which are associated with the rotational response of a quantum droplet. It would be 
interesting to confirm these results experimentally in this new superfluid system.

\end{document}